\documentclass[fleqn,leqno]{article}
\def\bfp#1{(\textbf{#1})}
\def\itp#1{(\textit{#1}\/)}
\newtheorem{lemma}{Lemma}
\newtheorem{theorem}{Theorem}
\def\sqr{{\vcenter{\vbox{\hrule height.4pt%
	\hbox{\vrule width.4pt height5pt \kern5pt%
	\vrule width.4pt} \hrule height.4pt}}}}
\newcommand{\qed}{\hfill$\sqr$}
\newenvironment{proof}{\begin{trivlist}
\item[\hspace{\labelsep}{\bf\noindent Proof: }]
}{\hfill\qed\end{trivlist}}
\newsavebox{\boxtabbing}
\newenvironment{boxtab}{\begin{lrbox}{\boxtabbing}%
\begin{minipage}{\columnwidth}\begin{tabbing}}%
{\end{tabbing}\end{minipage}\end{lrbox}%
\framebox[\columnwidth][l]{\usebox{\boxtabbing}}}

\newcommand{\remove}[1]{}
\renewcommand{\qed}{\hfill\rule{2mm}{2mm}}
\newtheorem{requirement}{Requirement}
\newtheorem{definition}{Definition}
\newtheorem{design}{Design}
\newenvironment{proofof}[1]{\begin{trivlist}
\item[\hspace{\labelsep}{\bf\noindent Proof of #1:}]
}{\hfill\qed\end{trivlist}}
\def\pgmv{\textsf}
\def\alt{\fbox{\raisebox{1ex}{\hspace*{-0.1em}\vspace*{1ex}}}}

\begin{document}
\title{Phase Clocks for Transient Fault Repair}
\author{Ted Herman\thanks{This work is supported 
by NSF CAREER award CCR-9733541.} \\
University of Iowa, Department of Computer Science \\
\texttt{herman@cs.uiowa.edu}}
\date{15 July 1999, revised 10 July 2000}
\maketitle
\begin{abstract} \noindent
Phase clocks are synchronization tools that 
implement a form of logical time in distributed
systems.  For systems tolerating transient faults by self-repair
of damaged data, phase clocks can enable reasoning about the 
progress of distributed repair procedures.  This paper presents
a phase clock algorithm suited to the model of transient memory
faults in asynchronous systems with read/write registers.   
The algorithm is self-stabilizing and guarantees 
accuracy of phase clocks within $O(k)$ time following an initial
state that is $k$-faulty.    
\end{abstract}
\section{Introduction} \label{introduction}

Measuring time is widely recognized as an important
system service and greatly simplifies the construction of many
distributed algorithms.  The reason, simply put, is that deductions
about the progress of concurrent activities, made by measuring elapsed
time, effectively substitute for communication and protocols that
directly monitor such progress.  Of course this technique can only be used
to the extent that a distributed system is synchronous, matching 
its progress with elapsed time.  Yet so attractive is the use of
time to simplify algorithm construction, that even in asynchronous
systems, researchers seek to simulate synchrony \cite{A85}, 
introduce logical clocks \cite{Lam78}, 
and/or logical time \cite{Mat89c} as programming tools.

One illustration of logical time in an asynchronous system is
the organization of a computation into phases.  
The basic property of a phased computation is that a process does
not enter phase $(k+1)$ until each related process has
completed phase $k$.  The case where all
processes are related is equivalent to barrier synchronization, 
and the case where the relation between processes is specified by
a graph corresponds to a \emph{phase clock}.   Many implementations of 
phased computation simply use a counter, called a \emph{clock}, to represent 
the current phase number of a process.  Consider the graph relation 
between processes to be a network communication topology, where the
graph has diameter ${\cal D}$ and distance between processes $p$ and
$q$ is denoted by $\textit{dist}_{pq}$.  A phase clock
invariantly relates phase numbers and distance as follows: 
any process $p$ has $\textit{clock}_p=k+d$ only if 
$\textit{clock}_q\geq k$ holds for each process $q$ 
satisfying $\textit{dist}_{pq}=d$ (notice that $k=1$ is just
the basic property mentioned above).   Thus if $\textit{clock}_p=k+d$,
holds at some state, we deduce that $\textit{clock}_q=k$ holds 
currently or held at some previous state.  This is a useful timing property 
because programs can use phase clocks for inferences about 
nonlocal information relayed through neighboring processes.
For example, process $p$ could use its clock to infer 
termination of a broadcast operation, rather than use 
explicit termination detection, by waiting for sufficiently many 
increments to $\textit{clock}_p$ (assuming that the broadcast
operation is geared to the phase clock).  

Phased computation is a reasonable discipline for many activities
of a distributed system, including procedures invoked as part of 
fault diagnosis and repair.  The fault domain for this paper is 
the model of \emph{transient faults}, which corrupt local process
states and communication registers, but do not damage a system's 
control logic.  It is therefore feasible for a system to self-diagnose
and restore variables corrupted by a transient fault to values
that enable correct system function.  One of the difficulties in 
using phase clocks to control distributed repair activities is 
that faults may corrupt local clock values.  The phase clock 
protocol presented in this paper not only repairs clock values 
corrupted by a transient fault, but does so in a manner that 
enables the system to use the phase clock for other repair 
activities.  

\paragraph{Contributions.}  This paper presents a 
distributed phase clock, called the \emph{repair timer}, 
specialized for the task of transient fault repair in a
distributed system.  The repair timer is 
\emph{time adaptive}, meaning that it satisfies desired
accuracy and progress properties within $O(k)$ time after
any transient fault event corrupting $k$ processes.   
The repair timer differs from standard phase clocks 
because it starts at zero and halts
after repair is complete (behaving somewhat like an egg-timer);  
this enables direct inspection of elapsed repair time, which standard phase
clocks do not provide\footnote{To see why this is not trivial,
suppose some time-adaptive phase clock were available, and 
consider measuring elapsed repair time by recording the 
start time of repair in some local variable; such a local
variable could, however, have an erroneous value due
to a transient fault.  Since transient
faults do not provide any signal at the start of repair, 
a process cannot locally decide whether its local variables are
accurate or not.}. The repair timer is also a 
self-stabilizing algorithm, able to restore all variables 
to a legitimate state following any transient fault event 
or combination of transient failures.  Finally, the paper presents 
composition theorems to show how the repair timer is useful
for the timing of fault repair procedures in a distributed
system. 

\paragraph{Related work.}  
Many recent works are motivated by what is seen
as pessimism in the model of self-stabilization, which does not
discriminate between cases of severe transient faults and 
minor transient faults.  In addition to the desired robustness
of self-stabilization, fast stabilization
of output variables has been recently demonstrated 
in a number of algorithms \cite{GGP96,GG96,BGK99} and some general methods 
to achieve time adaptivity \cite{KS97,KP98,DH99} 
or local self-stabilization \cite{GGHP96,AD97a}. 

Self-stabilizing phase clocks are given in \cite{GH90,ADG91,CFG92}.   
None of these constructions guarantee fast stabilization 
for cases of limited transient faults, and all appear to
require lengthy stabilization time (proportional to the
diameter of the communication graph) in some cases where
only a single process variable is corrupted by a transient fault. 
Requirements for a repair timer are described in \cite{H98b},
which is a precursor to this paper.

\paragraph{Contents.}  Section \ref{model} presents the computation
and system model for the paper.  Section \ref{algorithm} presents the
algorithm for the repair timer and Section \ref{properties} verifies 
the self-stabilization and time adaptive properties of the algorithm.
To illustrate the use of the repair timer, Section \ref{embedded} 
describes two designs incorporating the repair timer as a 
component in a system.  The paper's concluding remarks are
the subject of Section \ref{conclusion}.  Proofs of technical lemmas 
have been moved to the paper's Appendix.

\section{Distributed System} \label{model}

The system consists of a fixed set of $n$ processes that
communicate by reading and writing shared registers.
Communication between processes is limited to 
a network represented by an undirected, connected graph:  
for any pair of processes $(p,q)$
there exist a pair $(\pgmv{Register}_{pq},\pgmv{Register}_{qp})$ 
if and only if there is an edge between $p$ and $q$ in the 
communication graph.  Process $p$ is the only writer  
of $\pgmv{Register}_{pq}$ and $q$ is the only reader of 
$\pgmv{Register}_{pq}$.  A process cannot read the registers 
it writes.  Registers thus 
approximate message passing with bounded buffers, and 
a self-stabilizing simulation of link registers using messages
is described in \cite{D00}.  
A register can have numerous fields used to write values of  
different local variables (just as numerous local 
values can be transmitted in fields of one message).
\par
If $(p,q)$ is an edge 
in the communication graph, then $p$ and $q$ are called \emph{neighbors},
which is denoted by $p\in{\cal N}_q$ or equivalently, $q\in{\cal N}_p$.  
The diameter of the communication graph is $\cal D$. 
The distance between any pair $(p,q)$ in the graph is 
denoted by $\textit{dist}_{pq}$.  The term \emph{region} 
refers to a connected component of the graph
that has some property of interest.
\par
Each process is an autonomous, finite-state computing   
entity.  We use conventional imperative programming notation 
and concepts to describe the operation of a process, so 
each process has a program counter and program statements        
that manipulate variables.  A subset of these variables are
called \emph{output variables}, which directly support 
the system's intended function. 
\par
A \emph{configuration of p} is a specification of values, one for each of
process $p$'s variables, the value of $p$'s program counter, 
and a value for each register that $p$ writes.  
A (system) \emph{state} is a vector of process configurations, 
one configuration for each process in the system.
Any function from the set 
of all states to the set $\{\textit{true},\;\textit{false}\}$ 
is called a \emph{state predicate}.  
\par
A \emph{process step} is either a register operation (and corresponding 
advancement of the program counter) or some modification of 
internal and output variables (and program counter) of that process.  
A \emph{computation} is an infinite sequence of states 
so that each consecutive pair
of states corresponds to a process step and 
the sequence of states includes an infinite number
of steps of each process.  We thus assume that computations
are fair; more precisely, we assume weak fairness in that 
no process is prevented from executing steps in a computation. 
We use the term \emph{computation segment} to denote a finite,
contiguous subsequence of a computation.
\par
The program of each process specifies a \emph{cycle}, 
which consists of three parts:  \itp{i} a process reads 
the registers written by each of its neighbors, 
\itp{ii} the process possibly assigns values to its 
variables, and \itp{iii} the process writes registers for each
of its neighbors.  The definition of a cycle is 
a convenient and simple abstraction for measuring the progress
of a process in a computation. 
\par
The system is designed to accomplish some task represented
by a state predicate ${\cal L}_O$.  Whether or not ${\cal L}_O$ holds
at a given state is solely determined by the values of output
variables.   A predicate ${\cal L}$ is called a \emph{legitimacy
predicate} iff ${\cal L}$ is a system invariant and 
${\cal L}\Rightarrow{\cal L}_O$.  A state $\sigma$ is          
\emph{output-legitimate} if ${\cal L}_O$ holds at $\sigma$,
and is \emph{legitimate} if ${\cal L}$ holds at $\sigma$.    
It is often preferable to specify legitimacy 
(or output legitimacy) in terms of the behavior of processes rather
than explicitly specifying a state predicate.  A formal definition
of legitimacy in terms of behaviors is possible, but to streamline
the presentation, the state-based definition is used in this paper.  
Where process behavior is important in this paper, we verify 
separately that the system exhibits the desired behavior. 
\par
Because each iteration of a process program specifies a cycle,
time is conveniently measured in asynchronous rounds, which are
defined inductively.  A \emph{round} of a computation, with respect
to an initial state $\sigma$, is a computation segment originating
with $\sigma$ of minimum length containing at least one  
complete cycle (from reading registers to writing registers)    
of each process.  The \emph{first round} of a computation consists of 
a round with respect to the initial state of the computation,   
and \emph{round k} of a computation, $k>1$, consists of a round
with respect to the first state following round $k-1$.
\par
A round is, roughly speaking, one unit of ``parallel time'' in
the system.  A notion similar to a round is commonly used to analyze
the complexity of message-passing protocols by normalizing message
delay to the maximum message delay \cite{AW98}.  For analysis in 
this paper, the notion of a round is further refined.  An 
$\pgmv{R}_p^d$-round starting from a state $\sigma$ is a computation
segment of minimal length containing at least one complete
cycle of each process in the set $\{\,q\;|\;\textit{dist}_{pq}\leq d\;\}$.  
A round is thus equivalent to an $\pgmv{R}_p^{\cal D}$-round for
any choice of $p$.
\par
A system is \emph{self-stabilizing} if every computation contains
a legitimate state (that is, for any initial state, the system eventually 
reaches a legitimate state).  The \emph{stabilization time} is the 
worst case number of rounds in the prefix of a computation that 
does not contain a legitimate state.  Proving that a system is 
self-stabilizing entails demonstrating that a predicate $\cal L$
is invariant, implies ${\cal L}_O$, and that every computation 
contains some state satisfying $\cal L$.
\par
A fault event is a non-computational
operation that modifies variables, program counters, and/or registers.
More formally, a \emph{fault event} can be any pair of states  
(whereas a consecutive pair of states in a computation is a process step).
Computations do not include fault events;  a system history could
be a sequence of states consisting of computation segments
punctuated by fault events.  Reasoning about fault repair proceeds with
respect to each computation segment, since the system cannot 
anticipate whether or not another fault will occur.     
\par    
A state $\sigma$ is \emph{k-faulty} if $k$ is the minimum  
number of process configurations in $\sigma$ that, if appropriately
changed, transform $\sigma$ into a legitimate state.  The number
$k$ thus corresponds to the Hamming distance from $\sigma$ to
the nearest legitimate state.  There may be numerous ways
to transform $\sigma$ to a legitimate state by changing $k$ 
process configurations, some in which process $p$'s configuration
changes, and others where the transformation does not change
$p$'s configuration.  It is convenient to resolve this ambiguity
by some unique, deterministic choice of which processes should change
configurations to obtain a legitimate state from $k$-faulty state $\sigma$.
With such a deterministic choice, process configurations of $\sigma$
can be labelled \emph{faulty} or \emph{nonfaulty} depending on whether
they should change or not.  This deterministic choice can further be
refined to label variables and register fields as either faulty or 
nonfaulty.  How such a deterministic choice should be implemented
turns out not to be an issue in the sequel;  for the repair timer
given in Section \ref{algorithm} there is an unambiguous definition
of a faulty process configuration and for the interface proposed in
Section \ref{embedded} it is only required that if a faulty process
configuration neighbors a nonfaulty process configuration, then the
presence of a fault can be detected (which for many systems is the 
case even by reversing the choice of which of these two neighboring
configurations is considered to be faulty).  
\par
The main emphasis of this paper is time-adaptive, stable repair of output
variables, meaning that a system should stabilize its output variables
to satisfy ${\cal L}_O$ from any $k$-faulty initial state after at
most $O(k)$ rounds.  Formally, a system is \emph{time adaptive} if 
each computation starting from any $k$-faulty initial state $\sigma$  
contains an output-legitimate state $\sigma'$, within $O(k)$ rounds
following $\sigma$, such that every state following $\sigma'$ in
the computation is output-legitimate.  Given this emphasis, it is 
convenient to extend the terminology for faults:
a process $p$ is faulty (nonfaulty) in a computation iff
$p$'s configuration is faulty (nonfaulty) at the initial state.

\section{Algorithm} \label{algorithm} 

One of the difficulties in 
using phase clocks to control distributed repair activities is 
that faults may corrupt local clock values.  Indeed,
repair of the clocks values is a primary concern of this paper, and 
the usual timing properties of phase clocks must be modified to cope 
with faults.  Two goals for such modifications are:  \itp{a} clock values
of processes not affected by faults can reliably be used for inferences
about nonlocal information;  \itp{b} the response effort of the 
system is proportional to the scope of the fault.  

Goal \itp{a} seems relatively simple to satisfy, since the 
clocks of nonfaulty processes have predictable values.  However for
a standard phase clock, there are ambiguous cases of faulty situations.  
Suppose neighboring clocks have values $x$ and $x-2$ and only one of
these two is a faulty value;  there is no obvious way of distinguishing 
which of these two is faulty.  The approach taken in this paper is 
to use a specialized phase clock for fault repair called a 
\emph{timer}.   Whereas phase clocks advance throughout system
computation, the timer stops advancing when repair is complete.
Thus each timer clock reaches a prescribed value $\cal T$ 
when the system state is fully repaired.  If neighboring clocks
have values $\cal T$ and ${\cal T}-2$, then we may conclude that
the value ${\cal T}-2$ is due to a fault.

\begin{figure}[htb]
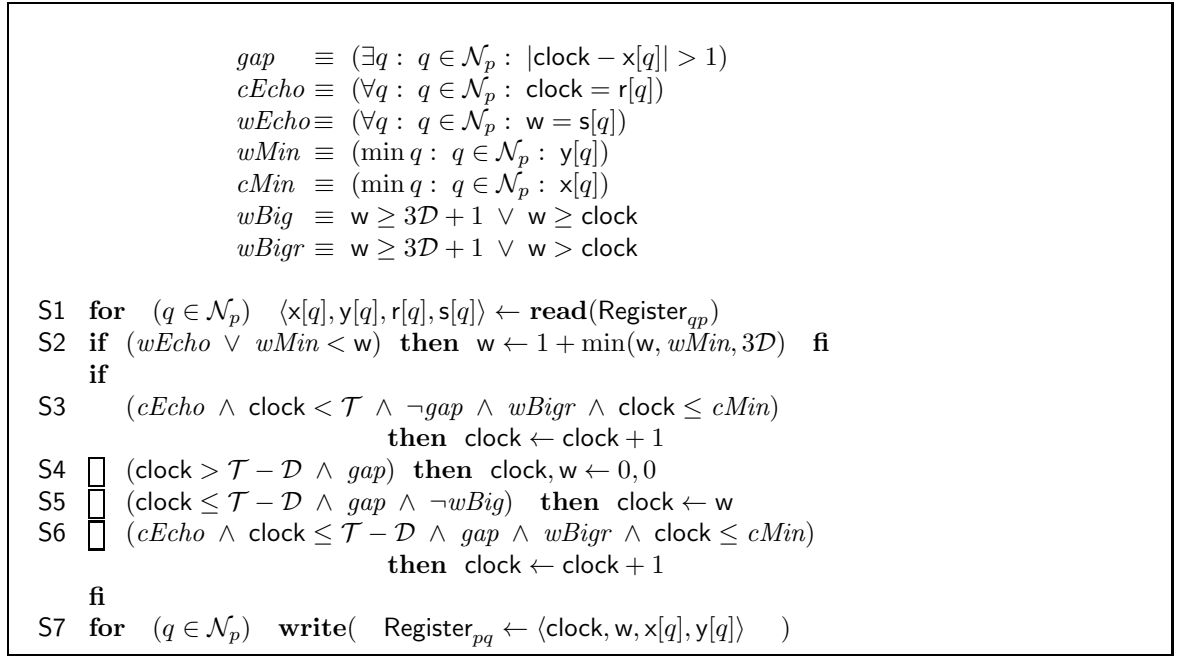

\begin{boxtab}
x \= xxx \= xx \= xx \= xx \= xx \= xxxxx \= xx \= xx \= \kill \\
\>\>\>\>\>\>
\textit{gap} \> $\equiv$ \> $(\exists q: \; q\in{\cal N}_p: 
	\; |\pgmv{clock}-\pgmv{x}[q]|>1)$ \\
\>\>\>\>\>\>
\textit{cEcho} \> $\equiv$ \> $(\forall q: \; q\in{\cal N}_p: \; 
	\pgmv{clock}=\pgmv{r}[q])$ \\
\>\>\>\>\>\>
\textit{wEcho} \> $\equiv$ \> $(\forall q: \; q\in{\cal N}_p: \; 
	\pgmv{w} = \pgmv{s}[q])$ \\ 
\>\>\>\>\>\>
\textit{wMin} \> $\equiv$ \> $(\min q: \; q\in{\cal N}_p: \; \pgmv{y}[q])$ \\
\>\>\>\>\>\>
\textit{cMin} \> $\equiv$ \> $(\min q: \; q\in{\cal N}_p: \; \pgmv{x}[q])$ \\ 
\>\>\>\>\>\>
\textit{wBig} \> $\equiv$ \> $\pgmv{w}\geq 3{\cal D}+1 \;\vee\; 
		\pgmv{w}\geq\pgmv{clock}$  \\ 
\>\>\>\>\>\>
\textit{wBigr} \> $\equiv$ \> $\pgmv{w}\geq 3{\cal D}+1 \;\vee\; 
		\pgmv{w}>\pgmv{clock}$  \\ \\ 
\> \pgmv{S1} \> \textbf{for} ~ ($q\in{\cal N}_p$) ~ 
        $\langle\pgmv{x}[q],\pgmv{y}[q],\pgmv{r}[q],\pgmv{s}[q]\rangle 
	\leftarrow$ \textbf{read}($\pgmv{Register}_{qp}$) \\ 
\> \pgmv{S2} \> \textbf{if} \> $(\textit{wEcho} \;\vee\; 
	\textit{wMin}<\pgmv{w})$ ~\textbf{then}~
	$\pgmv{w} \leftarrow 1+\min(\pgmv{w},\textit{wMin},3{\cal D})$ ~
	\textbf{fi} \\ 
\>\> \textbf{if} \\
\> \pgmv{S3} \>\> 
	$(\textit{cEcho}\;\wedge\;
	\pgmv{clock}<{\cal T}\;\wedge\;
	\neg\textit{gap} \;\wedge\;
	\textit{wBigr} \;\wedge\;
	\pgmv{clock}\leq \textit{cMin})$ \\
\>\>\>\>\>\>\>\>\> \textbf{then}~
	$\pgmv{clock}\leftarrow\pgmv{clock}+1$ \\
\> \pgmv{S4} \> \alt \> 
	$(\pgmv{clock}>{\cal T}-{\cal D}\;\wedge\;\textit{gap})$
	~\textbf{then}~
	$\pgmv{clock},\pgmv{w}\leftarrow 0,0$ \\
\> \pgmv{S5} \> \alt \> 
        $(\pgmv{clock}\leq{\cal T}-{\cal D}\;\wedge\;\textit{gap}
	\;\wedge\;\neg\textit{wBig})$ ~ \textbf{then}~
	$\pgmv{clock}\leftarrow \pgmv{w}$ \\
\> \pgmv{S6} \> \alt \> 
        $(\textit{cEcho} \;\wedge\;
	\pgmv{clock}\leq{\cal T}-{\cal D}\;\wedge\;
	\textit{gap}\;\wedge\;\textit{wBigr}\;\wedge\;
	\pgmv{clock}\leq\textit{cMin})$ \\
\>\>\>\>\>\>\>\>\> \textbf{then}~
	$\pgmv{clock}\leftarrow\pgmv{clock}+1$ \\
\>\> \textbf{fi} \\
\> \pgmv{S7} \> \textbf{for} ~ ($q\in{\cal N}_p$) ~ 
        \textbf{write}(~~ $\pgmv{Register}_{pq} \leftarrow 
	\langle\pgmv{clock},\pgmv{w},
        \pgmv{x}[q],\pgmv{y}[q]\rangle$ ~~ ) 
\end{boxtab}
\caption{timer for process $p$}
\label{clock}
\end{figure}\par
\paragraph{Variable Conventions.} 
The variables appearing in Figure~\ref{clock} are local  
variables of process $p$.  A number of proof arguments 
are statements relating variables of different processes, 
and subscripts are used to distinguish variable ownership 
(for instance, $\pgmv{clock}_q$ is owned by $q$).
Similarly, the predicates defined in Figure~\ref{clock}
are subscripted in definitions and proof arguments
(such as $\textit{gap}_p$ for process $p$).
\par
Statement \pgmv{S1} copies four register fields to 
four local variables, $\langle\pgmv{x,y,r,s}\rangle$.  Call these
variables the \emph{image} variables.  Implicitly
the code of Figure~\ref{clock} defines a mapping 
from each image variable to a register field and  
a corresponding ``base'' variable of a neighboring process
(written by statement \pgmv{S7}).  We say that    
each image variable is \emph{based on} a variable 
of a neighboring process, meaning that the value 
of an image variable is copied (via register communication) 
from the variable upon which it is based. 
Variable $\pgmv{x}_p[q]$, for example, is based on $\pgmv{clock}_q$.
Register fields are also images that are statically based
on variables.
\par
The meaning of time adaptivity described in Section \ref{model}
depends on declaring some of the process variables to be 
output variables.  For the repair timer, let $\pgmv{clock}_p$
be the output variable of process $p$.  The output correctness
for \pgmv{clock} variables is the subject of Section \ref{adaptivity}.  

\paragraph{Program Conventions.}
The statements \pgmv{S1}--\pgmv{S7} given in Figure \ref{clock}
describe one complete cycle of the repair timer for process $p$.
Therefore, after executing \pgmv{S7}, process $p$ executes 
\pgmv{S1} to start the next cycle.  The group of statements
\pgmv{S3}--\pgmv{S6} constitute a multiway \textbf{if} statement;  in
any cycle, at most one of \pgmv{S3}--\pgmv{S6} are executed.

Statements \pgmv{S2}--\pgmv{S6} specify internal calculations
for process $p$, since they manipulate local variables.  In a 
computation, we suppose that each of these statements specifies
one computation step.  Statements \pgmv{S1} and \pgmv{S7} specify
each $|{\cal N}_p|$ computation steps, since a step can read or
write at most one register.  Ordering of the read and write operations
of \pgmv{S1} and \pgmv{S7} is unimportant to the algorithm.
 
\paragraph{Algorithm Structure.}
To understand the algorithm of Figure~\ref{clock} it is useful to
first ignore statements \pgmv{S3}--\pgmv{S6} and focus attention on
the \pgmv{w} variables.  Notice that statement \pgmv{S2} will reduce
$\pgmv{w}_p$ if any of the registers 
read by \pgmv{S1} imply a value $\pgmv{w}_p-2$
or smaller for any neighboring \pgmv{w} variable.  The global effect of
many processes executing \pgmv{S2} can thus be ``convergence to the least
\pgmv{w}'' over a number of rounds.  The result of executions of \pgmv{S2}
will, in general, lead to a situation where neighboring process \pgmv{w}
variables differ by at most 1, which is one of the properties of a 
phase clock.  The \pgmv{wEcho} condition of \pgmv{S2} allows any process
with a globally minimal \pgmv{w} variable to increment its \pgmv{w} 
variable after all neighbors acknowledge its current value, via the 
\pgmv{s} image variables (which occurs within two rounds).  Therefore
the set of \pgmv{w} variables apparently enjoy both properties of a 
phase clock --- that neighboring \pgmv{w} variables differ by at most
1 and increase continually (until the upper bound of $3{\cal D}+1$ occurs)
in a computation.  
\par
Why not simply use the \pgmv{w} variables for 
repair timing and dispense with the logic of \pgmv{S3}--\pgmv{S6}?
The answer lies in the additional constraint we impose for faulty
initial states.  For repair purposes, it is not enough for clocks to 
be in phase and increment, they should also be accurate, meaning that
the value of a clock should be a measure of how long computation has
progressed after the initial detection of a fault.  The \pgmv{w} variables
do not have this property.  For instance, a faulty initial state could
have $\cal D$ as the initially smallest \pgmv{w} variable, so that all    
subsequent states have \pgmv{w} variables overstating the repair
time at least by $\cal D$.  An attempt to fix this problem would be
some statement similar to \pgmv{S4}, that would reset \pgmv{w} to zero
whenever neighboring \pgmv{w} variables differ by more than 1.  It is 
easy to construct examples of computations where such an attempted fix
will fail because \pgmv{w} variables are reset to zero infinitely often.  
This kind of idea can work, however, if any \pgmv{w} variable were
guaranteed to be reset to zero at most once in a computation, and 
that is the basic idea behind statements \pgmv{S3}--\pgmv{S6}, which
reset a \pgmv{clock} variable to zero at most once in a computation.
Although \pgmv{w} variables do not enjoy the accuracy needed for 
repair timing, they provide a useful ``reset layer'' for the \pgmv{clock}
adjustments of the algorithm.
\begin{definition} \label{timfin} ~ \emph{
A state is \emph{timer-final} if every register field and
image variable value is equal to the value of its base variable, and
$\;(\forall p:: ~ \pgmv{clock}_p={\cal T} 
	\;\wedge\; \pgmv{w}_p=3{\cal D}+1)$.
A process configuration is \emph{timer-final} if all its variables
and register fields have values corresponding to a timer-final state.
We define predicate ${\cal L}_T$ to hold for a state iff that state is 
timer-final. 
}\end{definition}
The value $\cal T$ used in the algorithm and Definition 
\ref{timfin} is a constant adequate for the fault tolerance
of the repair timer and for the application of the timer,
as discussed in Section \ref{embedded}.  The proof of self-stabilization
of the repair timer requires only that  
${\cal T}\geq 11{\cal D}$.
\par
\paragraph{Verification.}  
The verification of desired repair timer 
properties is divided into two stages.  First, the algorithm is 
considered as an isolated component, so that faulty states are
those states deviating from Definition \ref{timfin}.  
Section \ref{stabilization}
is devoted to a proof that the repair timer self-stabilizes to
a timer-final state.  Section \ref{adaptivity} presents the proofs that 
apply to $k$-faulty initial states, showing that the repair timer
achieves desired accuracy after $O(k)$ time following the 
initial state.  

The second stage of verification is concerned with
integration of the repair timer as a component of a system.   
The timer is a tool for time adaptive repair.  Discussion of how
the timer is used is deferred to Section \ref{embedded}, where it is explained
that the timer is a service with only one operation, namely 
to start the timer by assigning $\pgmv{clock}\leftarrow 0$;  
thereafter, the \pgmv{clock} should increment as a phase clock.  
Although a system state's legitimacy depends on variables of all
system components, the simple interface between the timer and other
system components makes it reasonable to consider 
fault tolerance properties of the timer in isolation,  
which motivates the two stage approach to verification.

\section{Stabilization and Adaptivity} \label{properties}
\subsection{Self-Stabilization} \label{stabilization}

Each process writes its communication registers in every
cycle from its variables. Therefore, 
following the first round of any computation, 
all register fields are equal to current or  
previous values of the corresponding base variables.  
Following the second round, each image variable has a 
value previously written from the corresponding base variable.
Moreover, following the third round of a computation,  
the third and fourth fields of $\pgmv{Register}_{qp}$ contain
values previously written by $p$ and then 
copied by $q$.  It is convenient to assume that register fields
correspond to values previously written in the computation,
so we call a computation \emph{based} if it is the suffix,
starting from round three or higher, of another computation.
\par
Statements \pgmv{S4} and \pgmv{S5} have the  
only assignments that may reduce the value of \pgmv{clock}
variables.  We call a computation (or computation segment)
\emph{reset-free} if no process executes \pgmv{S4} or 
\pgmv{S5} in that computation.  A computation is called 
\emph{rising} if it is the suffix of a based, reset-free
computation such that each process has read its registers
at least once in the based, reset-free computation prior to
the first state of the suffix.  Rising computations enjoy
the useful property that at all states, the value
contained in $\pgmv{x}_p[q]$ is a lower bound on the 
current value of $\pgmv{clock}_q$. (This property follows
because the computation is reset-free and each process
previously read registers and assigned to its \pgmv{x}
variables while the computation was reset-free.)
\begin{definition} \label{c0} \emph{
\[ b_{pq} \equiv (p\not\in{\cal N}_q) \;\vee\;
	(|\pgmv{clock}_p-\pgmv{clock}_q|<2 \;\wedge\; 
	\pgmv{x}_p[q]\leq\pgmv{clock}_q \;\wedge\;
         |\pgmv{clock}_p-\pgmv{x}_p[q]|<2) \] 
A state is \emph{smooth} if $(\forall p,q:: \; b_{pq})$. 
A set of processes $P$ forms a \emph{smooth region} if 
the subgraph of the communication topology induced by 
$P$ is connected and $(\forall p,q:\; p,q\in P: \; b_{pq})$.
}\hfill\qed\end{definition}
\begin{lemma} \label{a2} \emph{
In a rising computation, 
$(b_{pq}\;\wedge\;b_{qp})$ is an invariant for
any pair of processes $p$ and $q$.
}\end{lemma}
\begin{lemma} \label{a3} \emph{
Let $\sigma$ be the first state of a rising computation segment
such that for $p\in{\cal N}_q$, both $\pgmv{clock}_p$ and 
$\pgmv{clock}_q$ have incremented at least once in the computation segment.
Then $(b_{pq} \;\wedge\; b_{qp})$ holds at state $\sigma$.
}\end{lemma}
\begin{lemma} \label{a4} \emph{
If each \pgmv{clock} variable has incremented at least once
prior to state $\sigma$ in a rising computation segment, then
$\sigma$ is smooth.
}\end{lemma}
\begin{lemma} \label{a1} \emph{ 
Smoothness is invariant for a based computation;  
within $O({\cal D})$ rounds following a smooth state, 
a based computation contains a timer-final state.
}\end{lemma}
\begin{lemma} \label{a5} \emph{
If $\pgmv{clock}_p$ is less than $\cal T$ and less than or equal to all
neighboring \pgmv{clock} values at the initial state
of a based, reset-free computation segment, and $\pgmv{w}_p=3{\cal D}+1$
holds at the initial state, and this computation
segment contains at least two rounds, then $\pgmv{clock}_p$ 
increments at least once in the computation segment.
}\end{lemma}
\begin{lemma} \label{a6} \emph{
Let the initial state of a based computation satisfy
$(\forall p::\;\pgmv{clock}_p\leq 7{\cal D}\;\wedge\;\pgmv{w}_p=3{\cal D}+1)$.  
The computation contains a state where 
$(\exists q::\;\pgmv{clock}_q=10{\cal D}+1)$;
the first state satisfying $(\exists q::\;\pgmv{clock}_q=10{\cal D}+1)$
is a smooth state. 
}\end{lemma}
\begin{lemma} \label{a7} \emph{
Let the initial state of a based computation satisfy
$(\forall p::\;\pgmv{clock}_p\leq 7{\cal D}\;\wedge\;\pgmv{w}_p=3{\cal D}+1)$.  
Within $O({\cal D})$ rounds, the computation contains a smooth state.
}\end{lemma}
\begin{lemma} \label{a8} \emph{
Consider a based computation such that 
$(\pgmv{clock}_r=0\;\wedge\;\pgmv{w}_r=0)$ 
holds for some process $r$ in the initial state.
Within $\cal D$ rounds there is a state satisfying
$(\forall p:: \; \pgmv{clock}_p\leq 3{\cal D} \;\wedge\;
\pgmv{w}_p\leq 3{\cal D})$. 
}\end{lemma}
\begin{lemma} \label{a9} ~ \emph{
Consider a based computation such that 
$(\pgmv{clock}_r=0\;\wedge\;\pgmv{w}_r=0)$ 
holds for some process $r$ in the initial state.
Within $O(\cal D)$ rounds there is a smooth state or 
there is a state satisfying $(\forall p:: \; 
\pgmv{clock}_p\leq 7{\cal D} \;\wedge\; \pgmv{w}_p=3{\cal D}+1)$.
}\end{lemma}
\begin{theorem} \label{stab} \emph{
The timer stabilizes to a timer-final state (satisfying ${\cal L}_T$) 
in $O(\cal D$) rounds. 
}\end{theorem}
\begin{proof}
The invariance of ${\cal L}_T$ is verified 
by observing that none of \pgmv{S1}--\pgmv{S6} 
change any variable value at a timer-final state.
Convergence is demonstrated by a sequence of claims about
an arbitrary computation $A$.  Let $B$ be a suffix of $A$
beginning following the second round of $A$; by definition,
$B$ is a based computation.  We consider two cases for $B$.

\textbf{Case:} $B$ contains no step
executing \pgmv{S4} within $O({\cal D})$ rounds.
By arguments similar to those
given in the proof of Lemma \ref{a9}, some state of 
$B$ satisfies $(\forall p:: \; w_p=3{\cal D}+1)$ within
$O({\cal D})$ rounds and continues to hold at least until
\pgmv{S4} executes.  Let $C$ be a suffix of $B$ satisfying  
$(\forall p:: \; w_p=3{\cal D}+1)$ at its initial state.
Observe that $C$ is based and reset-free for $O({\cal D})$ rounds,
so Lemma \ref{a5} is applicable to $C$.  Within $O(2{\cal T})$
rounds of $C$, $(\forall p:: \; \pgmv{clock}_p={\cal T})$ holds, 
and the state satisfies ${\cal L}_T$.

\textbf{Case:} $B$ contains some step executing \pgmv{S4} within
$O({\cal D})$ rounds.  Execution of \pgmv{S4} results in a state
satisfying the premise of Lemma \ref{a9}.  Therefore $B$ either 
contains a smooth state within $O({\cal D})$ rounds, or contains
a state satisfying $(\forall p:: \; 
\pgmv{clock}_p\leq 7{\cal D} \;\wedge\; \pgmv{w}_p=3{\cal D}+1)$
within $O({\cal D})$ rounds.  The latter possibility is 
the premise for Lemma \ref{a7}, which shows that a smooth state 
is subsequently obtained within an additional $O({\cal D})$ rounds,  
so with either possibility, $B$ contains a smooth state within
$O({\cal D})$ rounds.  Lemma \ref{a4} implies that $B$ contains
a timer-final state within $O({\cal D})$ rounds following a smooth
state.
\end{proof}

\subsection{Time Adaptivity} \label{adaptivity}

The desired fault tolerance of the timer 
consists, informally, of the following two properties. 
\bfp{1} Within $k$ rounds from a $k$-faulty initial state, 
every \pgmv{clock} is accurate, that is, if $\pgmv{clock}_p=t$ 
for $t<{\cal T}$, it should be that $p$ has incremented 
$\pgmv{clock}_p$ as a phase clock $t$ times during the repair procedure.  
\bfp{2} Each faulty process \pgmv{clock} is reset to zero and
subsequently increments as a phase clock, incrementing to $k$ within
$O(k)$ rounds.

Property \bfp{1} provides the accuracy needed so that a process
can safely wait for distant information to be reliable.  Property \bfp{2}
assures that such distant information arrives in a timely fashion.  
Because faults may damage \pgmv{clock} and other timer variables, 
Theorem \ref{d6} below provides a 
conditional form of \bfp{1}, necessarily relaxed
to accommodate unusual initial states.  Also, some unusual       
cases of initial states require a conditional form of \bfp{2}, provided
by Theorem \ref{d7}.

A system state is faulty if it does not satisfy the definition of legitimacy. 
In considering the timer in isolation, a state is $k$-faulty if 
no fewer than $k$ process configurations require change to obtain
a timer-final state.  However, a complete definition of system legitimacy 
depends on components other than the timer, so a limited notion 
of fault is appropriate for the timer.  
\begin{definition} \label{perturb} ~ \emph{
A set of processes $P$ is \emph{unperturbed} at state $\sigma$ if 
$P$ forms a smooth region, 
$(\forall p: \; p\in P: \; \pgmv{clock}_p>{\cal T}-{\cal D})$, and
$(\forall p,q: \; p\in P\;\wedge\;q\in{\cal N}_p\;\wedge\;q\not\in P:\;
\pgmv{clock}_p={\cal T}\;\wedge\;\pgmv{x}_p[q]\geq{\cal T}-1)$.
A process $p$ is \emph{unperturbed} at $\sigma$ if 
there exists an unperturbed region containing $p$; process $p$ is 
\emph{perturbed} if there exists no unperturbed region containing $p$.
State $\sigma$ is $k$-\emph{perturbed} iff $k$ is the number of
perturbed processes at $\sigma$.  
}\hfill\qed\end{definition}
The motivation for this definition derives from 
the ambiguity of certain \pgmv{clock} values and nondeterminism of 
asynchronous computation.  Some proofs are simplified using Definition 
\ref{perturb}, which defines a perturbed process to be
a weakening of a faulty process configuration (a nonfaulty process 
configuration is unperturbed, but the converse may not hold).  It follows
that if the timer algorithm satisfies desired properties \bfp{1}--\bfp{2}
within $k$ rounds from any $k$-perturbed state, then similar properties
also hold for any $k$-faulty initial state.   Definition \ref{perturb} 
is not useful if $k=0$, so in the sequel any reference to $k$-perturbed
state is assumed to imply $k>0$.
\begin{definition} \label{accurate} ~ \emph{
Within a computation, a variable $\pgmv{clock}_p$ is 
\emph{d-accurate} at a state $\sigma$
if $\pgmv{clock}_p>{\cal T}-{\cal D}$ holds, or if 
$\pgmv{clock}_p\leq {\cal T}-{\cal D}$ implies, for
$0\leq m\leq{\cal D}$, that the number of 
$\pgmv{R}_p^m$-rounds completed prior to state $\sigma$
is at least $(\pgmv{clock}_p-m-d)$,
and that for every process
$q$, the value of $\pgmv{clock}_q$ has incremented at least 
$(\pgmv{clock}_p-\textit{dist}_{pq}-d)$ times prior to state
$\sigma$ in the computation.  For a computation 
initiating from a $k$-perturbed state, a state $\sigma$ is 
\emph{time-accurate} if for unperturbed $p$, 
$\pgmv{clock}_p$ is $d$-accurate for $d=2\cdot\min(k,{\cal D})$,
and for perturbed $p$, 
$\pgmv{clock}_p$ is $d$-accurate for $d=5\cdot\min(k,{\cal D})$.
}\hfill\qed\end{definition}
Definition \ref{accurate} falls short of the desired precision of 
property \bfp{1}, but satisfies safety concerns for many situation of 
repair timing because a $d$-accurate \pgmv{clock} provides a lower bound
on the number of cycles that distant processes have completed during
repair.  For instance, a repair application could depend on a distributed
procedure that terminates after $m$ clock increments in 
a non-faulty environment;  this application could wait for $d+m$ 
\pgmv{clock} increments if the repair timer ensures only $d$-accurate
\pgmv{clock} variables.  Unfortunately, an initially faulty state
can have arbitrary values in faulty process \pgmv{clock} variables, 
making it impossible to instantly have time accuracy.
Theorem \ref{d6} given at the end of this section states that 
time accuracy is guaranteed from any $k$-faulty initial state,
provided $k<n$, after at most $\min(k,{\cal D})$ rounds of computation. 
\begin{lemma} \label{d0} ~ \emph{        
Any process $p$ executes \pgmv{S4}, resetting 
$\pgmv{clock}_p$ and $\pgmv{w}_p$, at most once
in any computation.
}\end{lemma}
Lemma \ref{d0} is a corollary of arguments
given in the proofs of Lemmas \ref{a8} and \ref{a9}.
It is useful to know that processes execute \pgmv{S4}
at most once because any reset step subsequent to 
\pgmv{S4} is therefore due to \pgmv{S5}.  Arguments
in the proof of Lemma \ref{a9} show that 
\pgmv{w} values increase if \pgmv{S4} does not 
execute, and this idea can be used to establish
the eventual increase of \pgmv{clock} values.   
\begin{lemma} ~ \label{d2} \emph{
Let $\sigma$ be a result of $p$ executing \pgmv{S4}.  Then
for any process $q$ satisfying 
$\textit{dist}_{pq}=t\leq \min(k,{\cal D})$ there
occurs a state $\sigma'$, within $t$ rounds following $\sigma$,
such that $\pgmv{clock}_q\leq 3t \;\wedge\; \pgmv{w}_q\leq 3t$; 
and if there is a path consisting of unperturbed processes
from $p$ to $q$, then a state $\sigma''$ occurs within $t$ 
rounds following $\sigma$ 
such that $\pgmv{clock}_q\leq t \;\wedge\; \pgmv{w}_q\leq t$. 
}\end{lemma}
Lemma \ref{d2} considers a level of detail not discussed
in the proof of Lemma \ref{a8}, which supposes based computations.
Lemma \ref{d2} can also be extended to distances beyond $k$,
shown in the following.
\begin{lemma} ~ \label{d1} \emph{
In any computation starting from a $k$-perturbed initial state,
for each unperturbed process $p$ satisfying $\textit{dist}_{pq}=t$ 
with respect to some perturbed process $q$, the following holds: 
process $p$ executes \pgmv{S4} within $4+\min({\cal D},k+t)$ rounds.
}\end{lemma}
\begin{lemma} \label{d3} ~ \emph{
In any computation beginning from a $k$-perturbed state,
any process $p$ satisfying $\textit{dist}_{pq}=t$ from some
perturbed process $q$ does not execute \pgmv{S4} after 
round $4+\min(2{\cal D},t+2k)$. 
}\end{lemma}
\begin{lemma} \label{d4} ~ \emph{
Let $\sigma$ be a result of $p$ executing \pgmv{S4}.  Then
for any process $q$, within $t$ rounds following $\sigma$
there occurs a state $\sigma'$ such that        
$\pgmv{w}_q\geq \min(\lfloor(t-\textit{dist}_{pq})/2\rfloor,3{\cal D}+1)$ 
is invariant for the computation beginning with $\sigma'$.
}\end{lemma}
\begin{lemma} \label{d5} ~ \emph{
Let $\sigma$ be a result of $p$ executing \pgmv{S4}.  Then
for any process $q$, within $t+2$ rounds following $\sigma$
there occurs a state satisfying
$\pgmv{clock}_q\geq \min(\lfloor((t-2)-\textit{dist}_{pq})/2\rfloor,{\cal T})$. 
}\end{lemma}
\begin{theorem} \label{d6} ~ \emph{
Any computation starting from a $k$-perturbed initial state, $k<n$, 
contains a time-accurate state $\sigma$ after at most 
$\min(k,{\cal D})$ rounds following the
initial state, and all states following $\sigma$ are time-accurate states.
}\end{theorem}
\begin{proof}
Provided $k<n$, arguments in the proof of Lemma \ref{d1} show that
for each perturbed region $R$, some process $r$ 
executes \pgmv{S4} within the first round, where $r$
satisfies either $r\in R$ or $r\in {\cal N}_q$ for some $q\in R$. 
Lemma \ref{d2} then implies that within $\min(k,{\cal D})$
additional rounds, each $p\in R$ satisfies 
$\pgmv{clock}_p\leq 3\cdot\min(k,{\cal D})$.
Each unperturbed process \pgmv{clock} variable
remains larger than ${\cal T}-{\cal D}$ until
\pgmv{S4} is executed, which resets the \pgmv{clock}
to zero.  Thus within $\min(k,{\cal D})$ rounds, each
$\pgmv{clock}_p$ is either larger than ${\cal T}-{\cal D}$
or is at most $3\cdot\min(k,{\cal D})$.  After   
$\min(k,{\cal D})$ rounds, unperturbed processes can 
decrease \pgmv{clock} variables to zero, but such a 
decrease does not falsify the conditions for a time-accurate state. 
Therefore, to show time accuracy,
it suffices to show that increments to $\pgmv{clock}_p$ imply
corresponding increments have executed at distant processes. 

After a process $p$ executes \pgmv{S4}, it does not increment 
$\pgmv{clock}_p$ until \textit{cEcho} holds.  If an 
unperturbed $q$ is a neighbor of $p$, then $p$ does not 
increment $\pgmv{clock}_p$ until $q$ has reset $\pgmv{clock}_q$
and updated the image variables and register fields so that
$p$ observes \textit{cEcho}.  It is a simple induction to 
show that $\pgmv{clock}_p$ cannot increase to a value $t$ unless
$q$ has incremented $\pgmv{clock}_q$ at least $(t-1)$ times.  Now 
consider a minimum length path $P$ of processes, of length $d$,
from $p$ to some process $r$, such that each process in $P$ 
is unperturbed.  By a double induction, on $t$ and $d$, 
it follows that $\pgmv{clock}_p$ cannot increase from zero to
$t$ unless each process $q\in P$ has incremented 
$\pgmv{clock}_q$ at least $t-\textit{dist}_{pq}$ times.    
The same argument shows that processes
of $P$ complete at least the same number of $\pgmv{R}_p^d$-rounds
in the period where $\pgmv{clock}_p$ increases from zero to $t$.

Returning to the event of $p$ executing \pgmv{S4}, 
we now consider the case of perturbed $q\in{\cal N}_p$.  
As observed in the proof of Lemma \ref{d2}, it is possible
that $p$ can increment $\pgmv{clock}_p$ twice before 
$q$ completes a cycle because corrupt values in the 
initial state enable the \textit{cEcho} and \textit{wEcho}
conditions.  Furthermore, $p$ can increment $\pgmv{clock}_p$
a third time before $q$ increments its \pgmv{clock} because
$q$ completes a cycle to enable $\textit{cEcho}_p$.  However
in the case of such a third successive increment by $p$, 
$\pgmv{clock}_p>\pgmv{clock}_q$ and $b_{pq}\;\wedge\;b_{qp}$
hold as a consequence.  Thereafter, we reason about the
interaction between $p$ and $q$ as for unperturbed neighbors
(note that any subsequent executions of \pgmv{S5} by $p$ or 
$q$ validate this argument, since we reason about the 
highest value attained for \pgmv{clock} variables after
$p$'s initial three increments).  Therefore, the value of 
$\pgmv{clock}_p$ does not increase to $t$ unless $q$ 
has incremented $\pgmv{clock}_q$ at least $t-3$ times.    
Again, we may consider a minimum length path $P$ of processes, 
of length $d$, from $p$ to some process $r$, such that each process in $P$ 
is perturbed (with the possible exception of $p$).  
By a double induction, on $t$ and $d$, 
it follows that $\pgmv{clock}_p$ cannot increase from zero to
$t$ unless each process $q\in P$ has incremented 
$\pgmv{clock}_q$ at least $t-3\cdot\textit{dist}_{pq}$ times.    
Similar arguments show the completion of the appropriate
number of $\pgmv{R}_p^d$-rounds while $\pgmv{clock}_p$ increases
from zero to $t$.

Notice that in the case of a perturbed path of processes, accuracy
can diminish by two extra clock units per unit of distance, whereas
in the case of an unperturbed path, accuracy corresponds precisely
to distance.  These observations combined can be used to verify
that in any minimum length path $P$ from $p$ to $r$, after $p$
executes \pgmv{S4}, the value of $\pgmv{clock}_p$ increases 
to $t$ only if for each $q\in P$, the value of $\pgmv{clock}_q$ 
has incremented at least $t-\textit{dist}_{pq}-2m$ times, where
$m$ is the number of perturbed processes in the subpath of 
$P$ from $p$ to $q$.  Since $m\leq \min(k,{\cal D})$, time
accuracy is verified for $p$.  

The arguments above show that time accuracy holds for 
all unperturbed processes within $\min(k,{\cal D})$ rounds and
that any subsequent state is $2\cdot\min(k,{\cal D})$-accurate
for unperturbed processes.  For perturbed processes, similar
reasoning applies.  Instead of relying on \pgmv{S4} to establish
the baseline \pgmv{clock} value, we use instead a value bound
by the construction given in Lemma \ref{d2}'s proof.  Within
$\min(k,{\cal D})$ rounds, there is a state $\sigma'$ where
perturbed $p$ has a \pgmv{clock}
value of at most $3j$, and $j<\min(k,{\cal D})$ is the distance to some 
unperturbed process that executes \pgmv{S4} in the first round.  
The value of $\pgmv{clock}_p$ cannot increase from $3j$ to 
$3j+t$ unless process $q$ has incremented its \pgmv{clock}
at least $t-\textit{dist}_{pq}-2m$ times, where $m$ is at 
most $\min(k,{\cal D})$.  Therefore when $\pgmv{clock}_p=x$
at some state following $\sigma'$, we infer that 
$\pgmv{clock}_q$ has incremented at least 
$x-3\cdot\min(k,{\cal D})-\textit{dist}_{pq}-2\cdot\min(k,{\cal D})$
times, which verifies time accuracy for unperturbed processes.
\end{proof}
Theorem \ref{d6} addresses desired property \bfp{1} set out
at the beginning of the section.  Property \bfp{2} specifies
that each faulty process \pgmv{clock} be reset to zero and
then advance as a phase clock.  For the same reason that
\bfp{1} has been weakened to the time accuracy of 
Definition \ref{accurate}, we weaken \bfp{2} to require
only that each perturbed process be reset to some value
in the range $[0,3\cdot\min(k,{\cal D})]$ within $k$ 
rounds following the $k$-faulty initial state, and thereafter
increments as a phase clock.  Theorem \ref{d6} implies that
subsequent increases to \pgmv{clock} values satisfy 
a distance property relating the value of a \pgmv{clock}
to the number of increments of other \pgmv{clock} variables. 
The following theorem states the weakened form of \bfp{2}.
\begin{theorem} \label{d7} ~ \emph{
For any computation starting from a $k$-faulty initial state, $k<n$, 
each perturbed process \pgmv{clock} is at most $3\cdot\min(k,{\cal D})$
within $\min(k,{\cal D})$ rounds and increases to value 
$\lfloor((t-4)-\min(k,{\cal D}))/2\rfloor$ within $t$ rounds;
and each unperturbed process \pgmv{clock} 
similarly increases to $\lfloor(t-4)/2\rfloor$ within $t$ 
rounds after resetting by \pgmv{S4}. 
}\end{theorem}
\begin{proof}
Lemma \ref{d2} directly shows that perturbed processes assign
\pgmv{clock} variables to at most $3\cdot\min(k,{\cal D})$ within
the first $\min(k,{\cal D})$ rounds.  Lemma \ref{d5} establishes
that $p$ increases its \pgmv{clock} to at least  
$m=\min(\lfloor((t-2)-\textit{dist}_{pq})/2\rfloor$ after $t+2$
rounds following the execution of \pgmv{S4}.  Lemma \ref{d1}
establishes that for each perturbed region, 
some process $p$ either within or neighboring the perturbed
region executes \pgmv{S4} in the first round.  Lemma \ref{d5}
establishes that processes within a given distance increase
their \pgmv{clock} values as $\pgmv{clock}_p$ increases.  Any
process $q$ within a perturbed region containing or neighboring
$p$ is at most distance $\min(k,{\cal D})$ from $p$;  simplifying
the bound of Lemma \ref{d5} using $\min(k,{\cal D})$ as a distance
upper bound yields a lower bound of 
$\pgmv{clock}_q\geq\lfloor((t-2)-\min(k,{\cal D}))/2\rfloor$
after $t+2$ rounds.  
\end{proof}

\section{Embedded Timer} \label{embedded}

This section discusses use of the repair
timer as a component in a system.  Whereas   
Section \ref{properties} investigated properties of the repair timer in
isolation, the results of this section are essentially 
composition theorems stating conditions under which the 
repair timer can be used as a tool to enable time-adaptive 
fault tolerance in a system.  

Consider a system that uses the repair timer as one of its components.
The term \emph{core system} is used in this section to refer
to all system components outside the repair timer;  in other 
words, the entire system consists of the core system plus the repair
timer.  The elements of a process configuration (variables and registers)
can be partitioned into those belonging
to the repair timer and those belonging to the core system.    
The \emph{timer projection} of a state is formed by removing
all elements from each process configuration not relevant to the 
repair timer (that is, only \pgmv{clock}, \pgmv{w}, related image
variables and register fields are retained).  A \emph{core projection}
is formed by removing all repair timer elements from the state.  

\begin{requirement} \label{legit-composite} \emph{
Output legitimacy ${\cal L}_O$ 
of the system is defined solely in terms of
the core projection, that is, no repair timer variable is
an output variable.  Core system legitimacy, given by the predicate
${\cal L}_C$, is also defined with respect to the core projection;     
predicate ${\cal L}_C$ is independent of repair timer variables or 
register fields.  The legitimacy predicate for the system is
${\cal L}\equiv{\cal L}_C\;\wedge\;{\cal L}_T$.
\hfill\qed}\end{requirement}

\begin{figure}[htb]
\begin{picture}(400,100)(0,0)
\put(90,30){\framebox(80,30){core system}}
\put(250,30){\framebox(80,30){repair timer}}
\put(130,60){\line(0,1){5}}
\put(140,65){\oval(20,20)[lt]}
\put(140,75){\line(1,0){40}}
\put(185,72){$\pgmv{clock},\pgmv{w}\leftarrow 0,0$}
\put(265,75){\line(1,0){15}}
\put(280,65){\oval(20,20)[rt]}
\put(290,65){\vector(0,-1){5}}
\put(130,25){\vector(0,1){5}}
\put(140,25){\oval(20,20)[lb]}
\put(140,15){\line(1,0){40}}
\put(185,12){read \pgmv{clock}}
\put(235,15){\line(1,0){45}}
\put(290,25){\line(0,1){5}}
\put(280,25){\oval(20,20)[rb]}
\end{picture}
\caption{interface between core system and repair timer}
\label{interfig}
\end{figure}
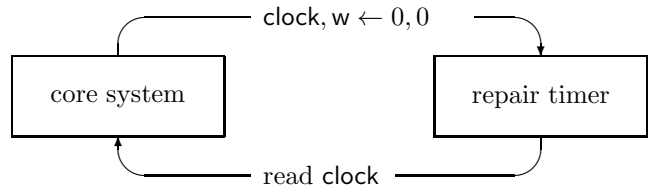

The interface between core system and repair timer is 
illustrated in Figure \ref{interfig}.  Communication between
these two components occurs in each process, but is restricted
to two methods:  the core system can reset the \pgmv{clock} and \pgmv{w}
variables, and the core system may read the current \pgmv{clock}
value.  Henceforth the term \emph{double-reset} is used to
denote the assignment $\pgmv{clock},\pgmv{w}\leftarrow 0,0$.  
Both \pgmv{S4}'s assignment of Figure \ref{clock} and the 
core system's assignment illustrated in Figure \ref{interfig}
are double-reset assignments.
 
So that results from Section \ref{properties} are applicable to 
the composite system, each process invokes 
the repair timer (statements \pgmv{S2}--\pgmv{S6}) 
once in each cycle.  Figure \ref{clock} includes \pgmv{S1} and \pgmv{S7} to 
present the repair timer in isolation, however in the context of 
a system invoking the repair timer, these two statements would be subsumed
by statements reading registers at the beginning of a process cycle
and writing registers at the end of a cycle.

A process configuration can be faulty with respect to 
the repair timer elements, the core system elements, or a combination
of both elements.  If a state $\sigma$'s core projection violates 
${\cal L}_C$ then $\sigma$ is said to be \emph{core-faulty};  if 
$\sigma$'s timer projection violates ${\cal L}_T$ then $\sigma$
is \emph{timer-faulty}.  While Definition \ref{timfin} provides the basis
for a precise characterization of a faulty repair timer, the 
situation for a general system can be ambiguous, as observed in
Section \ref{model}.  

\begin{requirement} \label{distinguish} \emph{
If $p$'s process configuration is not core-faulty
and $\pgmv{Register}_{qp}$ is faulty at $\sigma$,  
then the presence of a fault at $\sigma$ can be 
detected from the variables of $p$ and the contents of 
$\pgmv{Register}_{qp}$.
\hfill\qed}\end{requirement}

In many cases it is not difficult to design a system satisfying
Requirement \ref{distinguish}, in spite of the 
ambiguity of a faulty process configuration --- the requirement
only specifies that $p$ detect the \emph{presence} of a fault,
and $p$ is not required to determine the fault's location
(fault identification remains ambiguous).
Depending on the particular computation, 
$p$ may not detect a fault.  For instance,  
$q$ may repair its configuration, changing the contents
of $\pgmv{Register}_{pq}$, before $p$ reads the register. 

The importance of Requirement \ref{distinguish} is that 
nonfaulty $p$ has the capability to detect a fault, retain
the current values of its output variables, and initiate  
repair procedures.  Moreover, $p$ can ``contain'' the 
fault because it reacts before copying values from 
$\pgmv{Register}_{pq}$ and transmitting them to other processes. 

\begin{requirement} \label{interface} \emph{
Each cycle of a process invokes the repair timer.          
If, after reading registers at the start of a cycle,
a fault can be inferred (as described in Requirement \ref{distinguish})
for process $p$, and if $(\pgmv{clock}>{\cal T}-{\cal D})$,  
then $p$ executes a double-reset.
No other statements of the core system change the \pgmv{clock} or 
\pgmv{w} variables;  any number of statements of the core system
may read the \pgmv{clock} variable.  The legitimacy predicate
for the core system does not depend on the \pgmv{clock} or \pgmv{w} 
variables of the repair timer. 
\hfill\qed}\end{requirement}

\begin{requirement} \label{core-converge} \emph{
If any process $p$ executes a double-reset resulting
in a state $\sigma$, then within ${\cal T}-7{\cal D}$ rounds following
$\sigma$, the core system component of the state is legitimate.
\hfill\qed}\end{requirement}

Requirement \ref{core-converge} means, for most core systems,
that the core system stabilization time $\cal M$ satisfies 
${\cal M}\leq{\cal T}-7{\cal D}$.  In essence, this is a 
constraint on $\cal T$, which is added to the constraint
${\cal T}\geq 11{\cal D}$ given in Section \ref{algorithm}. 

\begin{lemma} \label{e1} \emph{
If the core system is self-stabilizing with stabilization time
$\cal M$ and satisfies Requirements 
\ref{legit-composite}--\ref{core-converge}, then the system 
is self-stabilizing with stabilization time ${\cal M}+O(\cal T)$,   
and the double-reset assignment executes 
at most once for each process in any computation.
}\end{lemma}

The proof of Lemma \ref{e1} rests on the independence of the core system
and the repair timer, as specified by Requirement \ref{interface}, and 
the fact that the core system stabilizes before there is any possibility
of executing a second double-reset by any process.  The requirements 
do not, however, preclude the design of the core system from depending
on repair timer properties.  For instance, proving stabilization
time $\cal M$ for the core system may depend on timer accuracy, 
since the core system can read \pgmv{clock} variables during 
convergence to ${\cal L}_C$, and timer accuracy 
can be used in some circumstances to measure the progress of distributed
algorithms and to allow processes to wait for such algorithms to stabilize. 

More interesting than using the repair timer for stabilization is the 
use of the repair timer to enable time adaptive repair of output variables.
The remainder of this section illustrates the use of the repair timer
in two designs.  Design \ref{construct-1} is a time adaptive
system, repairing output variables in $O(\min(k,{\cal D}))$ rounds 
from any $k$-faulty initial state.  The design requires that the 
core system use a sequence of repair procedures, following an idea
developed in \cite{DH99}.  Output variables of nonfaulty processes
may change to illegitimate values during convergence, but all output
variables satisfy ${\cal L}_O$ within $O(\min(k,{\cal D}))$ rounds
and continue to satisfy ${\cal L}_O$ thereafter.  
Design \ref{construct-2} is not fully time adaptive, but illustrates
another use of the repair timer:  the system can repair output variables
in $O(r)$ rounds from any $k$-faulty initial state, $k\leq r$, and no
nonfaulty process changes an output variable during repair.  

\begin{design} \label{construct-1} \emph{
The core system has ${\cal D}$ independent repair procedures, 
denoted $\pgmv{repair}^i$ for $1\leq i\leq {\cal D}$. 
Each of the repair procedures
uses its own set of variables, including variables that
are intended to be copied to the core system's output
variables.  Let $\pgmv{output}^i$ denote the set of variables
of $\pgmv{repair}^i$ that correspond to the system's output
variables, and let $\pgmv{repair}^i_p$ denote process $p$'s
portion of $\pgmv{repair}^i$.
We suppose that the core system also prepares a set of
variables $\pgmv{output}_C$ intended to be copied to output
variables. 
Each repair procedure is invoked in every process
cycle and $\pgmv{repair}^i$ is self-stabilizing to a predicate
${\cal L}^i$ within $\cal M$ rounds.  When ${\cal L}_C$ holds,
the $\pgmv{output}^i$ variables are equal to the system's output
variables, for $1\leq i\leq {\cal D}$, and $\pgmv{output}^C$
is also equal to the system's output. \\[1ex] 
Procedure $\pgmv{repair}^i$
has the property that, if the initial state is $j$-faulty, 
for $j\leq i$, then for all $p$, within $h\cdot i$ rounds, 
there occurs a state $\sigma$ such that for all $p$,  
the variables of $\pgmv{output}^i$ satisfy ${\cal L}_O$ (modulo
renaming or copying their values to the system outputs) at 
$\sigma$ and all subsequent states.  To exploit the repair 
timer, we suppose a stronger convergence property for
$\pgmv{repair}^i$, namely that $\pgmv{output}^i_p$ variables
stabilize within $h\cdot i$ of the $\pgmv{R}^{i}_p$-rounds. \\[1ex]
Given any $k$-faulty initial state satisfying $k\leq {\cal D}$, certain
\pgmv{repair} procedures agree on values for output variables:  
for $i\geq k$, after $\pgmv{repair}^i$ stabilizes $\pgmv{output}^i$, any 
procedure $\pgmv{repair}^{\ell}$ for $\ell>i$ stabilizes 
$\pgmv{output}^{\ell}$ to the same values that $\pgmv{output}^i$ has.     
Moreover, the core system stabilizes $\pgmv{output}^C$ 
to the same values contained in the stabilized $\pgmv{output}^k$ variables. 
The stabilized values of \pgmv{output} sets are also constrained
by distance from a fault:  for any nonfaulty $p$ such that the minimum
distance from $p$ to a faulty process is $d$, where $d>k$, then 
\emph{all} of the $\pgmv{output}_p$ sets stabilize to the same 
values already contained in $p$'s output variables.  \\[1ex]
The $\pgmv{output}_p$ sets are copied to the output variables of
process $p$ as follows.  In each cycle, if $\pgmv{clock}_p={\cal T}$, 
then $p$ copies $\pgmv{output}^C_p$ to its output variables.   
Otherwise, in each cycle, $p$ copies $\pgmv{output}^i_p$ to 
its output variables where $i$ is the largest value satisfying
$1\leq i\leq{\cal D}$ and $(h+5)\cdot i \leq \pgmv{clock}_p$.
No $\pgmv{output}_p$ set is copied to the output variables if 
$(h+5)\cdot{\cal D}<\pgmv{clock}_p<{\cal T}$ holds. 
\hfill\qed}\end{design}

\begin{theorem} \label{e2} \emph{
If a system for Design \ref{construct-1} satisfies Requirements
\ref{legit-composite}--\ref{core-converge} and  ${\cal M}=O({\cal D})$, 
then within $O(\min(k,{\cal D}))$ rounds following any $k$-faulty
initial state, the system output-stabilizes to ${\cal L}_O$.
}\end{theorem}

\begin{proof}
Consider a $k$-faulty initial state.  If $k\geq{\cal D}$, then 
from ${\cal M}=O({\cal D})$ and Theorem \ref{stab}, the system 
stabilizes to ${\cal L}$ and hence ${\cal L}_O$ in $O({\cal D})$
rounds, which proves the conclusion.  The remaining case is 
$k<{\cal D}$ for a $k$-faulty initial state.  For this case, 
we first show that any faulty process $\pgmv{clock}$ is 
time accurate within the first $k$ rounds.  Requirement
\ref{interface} ensures that some process within distance $k$ from any
faulty process executes a double-reset in the first round, and 
Theorem \ref{d6} implies subsequent time accuracy 
within $k$ rounds.  The same argument implies
that each nonfaulty process within distance $k$ from a faulty process
has a time-accurate \pgmv{clock} after at most 
$k$ rounds.  All nonfaulty processes have time-accurate 
\pgmv{clock} variables throughout the computation.    

Design \ref{construct-1} specifies that some nonfaulty processes
do not change their output variables by any \pgmv{repair} procedure,
so the proof obligation is to show that faulty processes and those
nonfaulty processes within distance $k$ to a faulty process stabilize
their output variables in $O(k)$ time.  After $k$ rounds, all such
processes have time-accurate \pgmv{clock} variables.  By definition
of time accuracy for a $k$-faulty initial state, a 
time-accurate $\pgmv{clock}_p$ variable with value $t$ implies
that the number of $\pgmv{R}^k_p$-rounds preceding in the computation is
at least $t-5k$.  Procedure $\pgmv{repair}^k$ converges
within $h\cdot k$ of the $\pgmv{R}^k_p$-rounds, so after time accuracy
holds, a $\pgmv{clock}_p$ value of $h\cdot k + 5k = (h+5)\cdot k$
implies that variables of $\pgmv{output}_p^k$ can be copied
to $p$'s output variables.  The conditions of Design \ref{construct-1}
also justify copying $\pgmv{output}_p^j$ to the output variables
when $\pgmv{clock}_p\geq (h+5)\cdot j$ for $k<j<{\cal D}$.

Having established the safety of copying \pgmv{output} sets 
to the output variables, the remaining obligation is to show
that all such copying either completes within $O(k)$ rounds
or that any subsequent copying will not affect ${\cal L}_O$. 
Theorem \ref{d7} implies that all processes within distance
$k$ to a faulty process will, 
after time accuracy holds, increase their \pgmv{clock}
variables to $(h+5)\cdot k$ within $O(k)$ rounds and will not
subsequently decrease their \pgmv{clock} values below this value.
Therefore, within $O(k)$ rounds, all processes within distance
$k$ to a faulty process assign their output variables, while 
those processes further than distance $k$ from a fault do not
assign their output variables to falsify ${\cal L}_O$ by any
step of the computation.   
\end{proof}

The $\pgmv{repair}^i$ procedures of Design \ref{construct-1}
are independent, meaning that they do not share any of the 
variables they modify.  Because the variables of $\pgmv{repair}^C$ 
are inactive for nonfaulty processes during the period of 
stabilization, they are a resource for faulty processes:  values
from nonfaulty $\pgmv{output}_p^C$ can be disseminated 
to other processes and used for the stabilization of 
$\pgmv{repair}^i$ procedures.
For details on this technique, illustrated in a synchronous 
computation model, the reader is referred to \cite{DH99}.

\begin{design} \label{construct-2} \emph{
The core system uses procedure $\pgmv{repair}^r$ with
a set of variables denoted $\pgmv{output}^r$ that 
are equal to the system output variables at a legitimate state.    
Procedure $\pgmv{repair}^r$ stabilizes the 
$\pgmv{output}^r$ variables to satisfy ${\cal L}_O$ 
within $h\cdot r$ time from any initial state that
is $j$-faulty for $j\leq r$;  each faulty 
process $p$ stabilizes $\pgmv{output}_p^r$ after 
at most $h\cdot r$ of the $\pgmv{R}_p^r$-rounds
occur.  The $\pgmv{output}^r_p$ sets are copied to output variables of
process $p$ as follows.  In each cycle, if $\pgmv{clock}_p\geq (h+5)\cdot r$, 
then $p$ copies $\pgmv{output}^r_p$
to its output variables;  for all other values of $\pgmv{clock}_p$
process $p$ leaves its output variables unchanged.  The $\pgmv{repair}^r$
procedure stabilizes $\pgmv{output}_p^r$ to values already contained
in $p$'s output variables for any nonfaulty $p$.
\hfill\qed}\end{design}

\begin{theorem} \label{e3} \emph{
If a system for Design \ref{construct-2} satisfies Requirements
\ref{legit-composite}--\ref{interface}, then within 
$O(r)$ rounds following any $k$-faulty initial state for $k\leq r$, 
the system output-stabilizes to ${\cal L}_O$.  No step
modifies output variables of nonfaulty processes to values
differing from those specified by ${\cal L}_O$. 
}\end{theorem}

Theorem \ref{e3} can be verified by reasoning similar to
the proof of Theorem \ref{e2}.  
Design \ref{construct-2} is not self-stabilizing and 
Theorem \ref{e3} does not  
specify Requirement \ref{core-converge} as a condition.  
The fault tolerance of this design is limited to $r$ faulty
processes.  

\section{Concluding Remarks} \label{conclusion}

It is challenging to construct a system that can repair variables
inflicted by transient faults.  A reasonable methodology for such
system construction is based on tools for fault detection and 
repair, and these tools must themselves satisfy properties of 
time adaptivity and stabilization.  This paper presented a phase
clock algorithm specialized for the task of fault repair.  The 
designs presented in Section \ref{embedded} show how the
repair timer can be composed with other system components. 
\par
Although time adaptivity and self-stabilization are major 
themes for this paper, the repair timer can be useful even
when neither full stabilization nor fast stabilization is needed,
because it is convenient to reason about the progress of repair
procedures by measuring elapsed time (which would otherwise be
complicated due to possible corruption of time-measurement variables). 
An observer of the system located at process $p$ could monitor 
repair progress by repeatedly examining $\pgmv{clock}_p$, possibly
delaying critical activity until repair is complete.  
\par
Use of the repair timer can add overhead to repair procedures
because each cycle of repair invokes the timer, and the \pgmv{clock}
variable only increments in relation to rounds.  It could be that
actual repair only involves a small subset of processes, but a 
\pgmv{clock} variable will not, in general, 
increment $t$ times unless \emph{all}
processes at distance $d$ have completed $t-d$ cycles --- including 
processes that are not involved in the repair.  Thus the measurement
of repair time in rounds could be overly pessimistic and cause
processes to wait longer than necessary before they infer that repair
is complete.  Another slowing of repair timing results if 
a loose upper bound on the network diameter is used for $\cal D$
(an upper bound is typically proposed for dynamic networks)
since $\cal T$, the ``resting value'' for the repair timer, is determined
by the value $\cal D$.

\section{Appendix:  Proofs}

\begin{proofof}{Lemma \ref{a2}}
The essence of the proof is that neither \pgmv{S3} nor \pgmv{S6}
increment a \pgmv{clock} to a value two greater than any neighbor.
Since Definition \ref{c0} involves \pgmv{x} variables, the     
effect of statement \pgmv{S1} requires examination. 
Reading a register to assign an \pgmv{x} variable only increases
the accuracy of the image variable;  in particular, given  
$(b_{pq}\;\wedge\;b_{qp})$ as a precondition, \pgmv{S1} does 
not falsify this condition, because $p$ and $q$ have \pgmv{clock}
values differing by at most one in the precondition.
Therefore it suffices to verify that 
any change to $\pgmv{clock}_p$ or $\pgmv{clock}_q$ also satisfies
the lemma.  In a reset-free computation, only \pgmv{S3} and \pgmv{S6} 
change a \pgmv{clock} variable.  If \pgmv{S3} executes, incrementing
$\pgmv{clock}_p$, we have $\pgmv{clock}_p\leq\pgmv{x}_p[q]$ as a
precondition.  Since $b_{pq}$, we have  
$\pgmv{clock}_p\in\{\pgmv{clock}_q,\pgmv{clock}_q+1\}$ also
as a precondition; thus the 
increment to $\pgmv{clock}_p$ results in a state satisfying
$|\pgmv{clock}_p-\pgmv{clock}_q|<2$, verifying $b_{pq}$.  
The postcondition also satisfies $b_{qp}$, since the change
to $\pgmv{clock}_p$ does not alter the relation between
$\pgmv{clock}_q$ and $\pgmv{x}_q[p]$.  A similar argument applies
to \pgmv{S6}, and also to the case of $q$ incrementing its \pgmv{clock}.
\end{proofof}
\begin{proofof}{Lemma \ref{a3}}
By definition of a rising computation, each process has a lower
bound on neighboring \pgmv{clock} variables in its \pgmv{x} variable,
because \pgmv{clock} values cannot decrease in a reset-free 
computation.  Suppose $p$ is the first of $(p,q)$ to increment
its \pgmv{clock}.  A precondition for this step is 
$\pgmv{clock}_p\leq\pgmv{x}_p[q]$, which implies 
$\pgmv{clock}_p\leq\pgmv{clock}_q$, which in turn implies
$\pgmv{x}_q[p]\leq\pgmv{clock}_q$.  Consider two cases for this
last inequality, \itp{i} $\pgmv{x}_q[p]<\pgmv{clock}_q$ or 
\itp{ii} $\pgmv{x}_q[p]=\pgmv{clock}_q$.  For \itp{i}, process $q$ 
cannot increment $\pgmv{clock}_q$, and this situation will persist
until $p$ increments its \pgmv{clock} sufficiently many times
so that $\pgmv{clock}_p\geq\pgmv{clock}_q$.  It is straightforward
to verify that $p$ does not increment $\pgmv{clock}_p$ beyond
$\pgmv{clock}_q+1$, so for case \itp{i} the first increment to 
$\pgmv{clock}_q$ establishes $(b_{pq} \;\wedge\; b_{qp})$.
For \itp{ii}, we deduce from the inequalities above
$(\pgmv{clock}_q\leq\pgmv{clock}_p\;\wedge\;
\pgmv{clock}_p\leq\pgmv{clock}_q)$ holds as precondition to 
$p$'s first increment step, and the inequalities with 
regard to the \pgmv{x} variables are similar.  So for case
\itp{ii},  $(b_{pq} \;\wedge\; b_{qp})$ holds directly. 
\end{proofof}
\begin{proofof}{Lemma \ref{a4}}
By Lemma \ref{a3}, neighboring $(p,q)$ establish 
$(b_{pq}\;\wedge\;b_{qp})$ at or before $\sigma$; 
by Lemma \ref{a2} such processes continue to 
satisfy this property for
the remainder of the reset-free computation segment.  
\end{proofof}
\begin{proofof}{Lemma \ref{a1}}
Because we consider a based computation, and not a rising
computation in this lemma, the invariance of 
$(b_{pq}\;\wedge\;b_{qp})$ stated in Lemma \ref{a2} is
not applicable.  Note that $(\forall p:: \; \neg\textit{gap}_p)$ 
holds at a smooth state. The invariance of smoothness 
is therefore verified from the conditions
of \pgmv{S3}--\pgmv{S6}, since no gap exists at a smooth state
and \pgmv{S3} preserves smoothness.   
It is also simple to verify that the least \pgmv{clock}
value, if smaller than $\cal T$, increments within two rounds
from a smooth state, hence at most $2{\cal T}=O({\cal D})$ rounds are 
needed to obtain a state satisfying $(\forall p:: \; 
\pgmv{clock}_p={\cal T})$.  A similar argument shows that
all \pgmv{w} variables converge to $3{\cal D}+1$ within $O(\cal D)$ rounds.
\end{proofof}
\begin{proofof}{Lemma \ref{a5}}
In the first round, $p$ reads neighboring \pgmv{clock} values 
and detects local minimality.  If $p$ increments in this round,
the lemma holds;  and if $p$ does not increment, it writes its
\pgmv{clock} and detects \textit{cEcho} in the next round, and 
local minimality implies $p$ will increment $\pgmv{clock}_p$ 
either by \pgmv{S3} or \pgmv{S6}.  
\end{proofof}
\begin{proofof}{Lemma \ref{a6}}
Observe that $(\forall p::\;\pgmv{w}_p=3{\cal D}+1)$ holds at least until some 
\pgmv{clock} exceeds ${\cal T}-{\cal D}$ so that \pgmv{S4} can 
execute.  And $\pgmv{w}_p=3{\cal D}+1\Rightarrow\textit{wBig}_p$, so
process $p$ does not execute the assignment of \pgmv{S5}.  
This implies that the computation is reset-free until 
some \pgmv{clock} obtains the value exceeding ${\cal T}-{\cal D}$.  
Lemma \ref{a5} implies that
each minimal \pgmv{clock} value increments in any pair of rounds,
which implies that the maximum of the set of \pgmv{clock} values
eventually grows as the computation proceeds.  Let $\pgmv{clock}_p$
be the first \pgmv{clock} to attain the value $8{\cal D}+1$ at state
$\alpha$. Thus $(\forall q:\; q\in{\cal N}_p: \; \pgmv{clock}_q\geq 8{\cal D})$   
holds prior to $\alpha$.  More generally, it follows by 
induction that $(\forall q:\; \pgmv{dist}_{qp}=k>0: \; 
\pgmv{clock}_q\geq 8{\cal D}-k)$.  Therefore, each \pgmv{clock} 
value has incremented at least once prior to $\alpha$.  Let $\pgmv{clock}_q$
be the first \pgmv{clock} to attain the value $9{\cal D}+1$ at state
$\beta$.  At state $\beta$, each process has incremented its \pgmv{clock}
twice in a reset-free computation, implying that each process has read all
of its registers at least once in this reset-free computation.  Therefore the 
computation segment beginning with $\beta$ is by definition a rising 
computation segment (at least until some \pgmv{clock} 
exceeds ${\cal T}-{\cal D}$).  Now let $\pgmv{clock}_r$
be the first \pgmv{clock} to attain the value $10{\cal D}+1$ at state
$\gamma$.  At state $\gamma$, each process has incremented its \pgmv{clock}
at least once in a rising computation, and by Lemma \ref{a4}, 
$\gamma$ is a smooth state. 
\end{proofof}
\begin{proofof}{Lemma \ref{a7}}
Lemma \ref{a6} shows that the computation contains a smooth state,
so the obligation here is to show the $O({\cal D})$ time bound.
By Lemma \ref{a5} each minimal clock value increments at least once in any
two consecutive rounds, so within $20{\cal D}+2$ rounds, some \pgmv{clock}
attains the value $10{\cal D}+1$, establishing a smooth state.
\end{proofof}
\begin{proofof}{Lemma \ref{a8}}
The proof begins with a claim on the first $k$ rounds 
of the based computation: within $k$ rounds there is 
a state satisfying 
\begin{eqnarray}
\label{a8A} ~ & ~ &  
(\forall q: ~ \textit{dist}_{qr}=k: ~ \pgmv{w}_q\leq 2k
\;\wedge\; \pgmv{clock}_q\leq 2k) \\
\label{a8B} ~ & ~ &	
(\forall q,j: ~  j<k \;\wedge\;\textit{dist}_{qr}=j: ~ 
\pgmv{w}_q\leq 3k-j \;\wedge\; \pgmv{clock}_q\leq 3k-j)
\end{eqnarray}
The claim is shown by induction.  The first state of the 
computation satisfies the claim for $k=0$ as the base case.  
Suppose the claim holds for $k\leq\ell$ and consider two 
processes $q$ and $s$ such that $\textit{dist}_{rq}=\ell$,
$s\in{\cal N}_q$, and $\textit{dist}_{rs}=\ell+1$.  Let 
$\sigma$ be a state satisfying (\ref{a8A})--(\ref{a8B}) for $k=\ell$.
By the \emph{cEcho} condition of \pgmv{S3}, process $q$ does
not increment $\pgmv{clock}_q$ beyond $2\ell$ until process
$s$'s image fields in $\pgmv{Register}_{sq}$ have the 
appropriate values.  In fact, these register fields may 
initially have the appropriate values, which would allow $q$ to increment
\pgmv{clock} and \pgmv{w} variables to $2\ell+1$ by \pgmv{S2}--\pgmv{S3}.  
However process $q$ cannot subsequently increment to $2\ell+2$
until the \emph{cEcho} condition holds, which requires a cycle
by $s$ (and all other neighbors).  Process $s$ therefore observes
$y_s[q]\leq2\ell+1$ in its cycle and assigns at most $2\ell+2$ to its 
\pgmv{w} and \pgmv{clock} variables.  Since $\sigma$ occurs at least by 
round $\ell$, the bound of $2\ell+2$ for $s$ variables applies 
within round $\ell+1$, which establishes (\ref{a8A}) of the claim.  
\par
Condition (\ref{a8B}) is also shown by induction.  For $k=0$, the base
case, (\ref{a8B}) holds vacuously.  Now suppose (\ref{a8B}) holds
for $k\leq\ell$ and consider two 
processes $q$ and $s$ such that $\textit{dist}_{rq}=\ell$,
$s\in{\cal N}_q$, and $\textit{dist}_{rs}=\ell+1$.  
Condition (\ref{a8A}) places an upper bound on variables at distance
$\ell+1$ from process $r$ within round $\ell+1$.  Therefore
$\pgmv{clock}_s\leq 2(\ell+1)$ within round $\ell+1$.  In moving from 
round $\ell$ to $\ell+1$, we consider the possibilities for 
process $q$ and $\pgmv{clock}_q$.  If $\pgmv{clock}_q$ and 
$\pgmv{clock}_s$ differ by more than one and process $q$ executes
a cycle, then \pgmv{S5} resets $\pgmv{clock}_q$;  before any 
further change to $\pgmv{clock}_q$ occurs, the \textit{cEcho} 
condition requires a full cycle by $s$, which validates (\ref{a8B})
up to distance $\ell+1$ within round $\ell+1$.  If $\pgmv{clock}_q$
and $\pgmv{clock}_s$ are equal or differ by one, then $\pgmv{clock}_q$
could increment.  Observe here that no \pgmv{clock} or \pgmv{w} variable 
can increment beyond one more than any neighboring
value;  by another inductive argument, 
no \pgmv{clock} or \pgmv{w} variable increments
beyond $j$ more than any corresponding variable at distance $j$.
Therefore $\pgmv{clock}_q$ does not increment beyond 
$(2\ell+2)+1$ so long as $\pgmv{clock}_s\leq 2\ell+2$.  
This observation is generalized by (\ref{a8B}) for $k=\ell+1$ within
round $\ell+1$.  Note that we have assumed that any \pgmv{clock}
increment is due to \pgmv{S3} and not \pgmv{S6} in this argument;  this
assumption is justified by (\ref{a8A}), since $\pgmv{w}<3{\cal D}+1$, which
disables execution of \pgmv{S6}.
\end{proofof}
\begin{proofof}{Lemma \ref{a9}}
Let $\sigma$ be a 
state satisfying $(\forall q:: \; \pgmv{clock}_q\leq 3{\cal D})$.
By Lemma \ref{a8} such a state $\sigma$ occurs 
with $\cal D$ rounds of the based computation.
So long as every \pgmv{clock} is at most ${\cal T}-{\cal D}$, no 
step subsequent to $\sigma$ decreases a \pgmv{w} variable;  and if
no \pgmv{w} variable is reset by \pgmv{S4} in a consecutive pair of rounds, 
then the minimum value of the set of \pgmv{w} variables either increases 
by that pair of rounds or all \pgmv{w} variables already have the 
maximum $3{\cal D}+1$ value (we consider a consecutive pair of rounds to 
ensure that \emph{wEcho} will hold for \pgmv{S2}).  
Therefore, if no \pgmv{clock} variable attains the value 
$7{\cal D}+1$ within $2\cdot(3{\cal D}+1)$ rounds, all
\pgmv{w} variables equal $3{\cal D}+1$ and the lemma holds.  On the other
hand, if some \pgmv{clock} does attain the value $7{\cal D}+1$, we shall 
deduce that all \pgmv{w} values equal $3{\cal D}+1$, which also proves
the lemma.  The argument rests on the following claim:  at all states
subsequent to $\sigma$ satisfying 
$(\forall p:: \; \pgmv{clock}_p\leq 7{\cal D})$, the 
implication $\pgmv{clock}_p\geq 3{\cal D}+k\Rightarrow\pgmv{w}_p\geq k$ 
holds for every $p$ and $0\leq k\leq 3{\cal D}+1$.  This claim is 
verified by induction on $k$.  For $k=0$ the result is immediate from
the domain of \pgmv{w} variables.  Now consider $k>0$ and suppose
the claim holds for $k-1$.  Let $q$ be the first process to assign
$\pgmv{clock}_q\leftarrow 3{\cal D}+k$.  If the assignment occurs by
\pgmv{S6} then $w=3{\cal D}+1$ and the claim holds;  if the assignment
occurs by \pgmv{S3}, then each neighbor of $q$ has a \pgmv{clock} value
of $3{\cal D}+(k-1)$, hence by hypothesis each neighboring \pgmv{w} 
variable is at least $k-1$, and $\pgmv{w}_q\geq k-1$ by the same
hypothesis.  The result is that the same cycle assigning 
$\pgmv{clock}_q\leftarrow 3{\cal D}+k$ also assigns $\pgmv{w}_q$ to 
be at least $k$.  Similar arguments treat the general case for 
$q$ (not necessarily the first) assigning $3{\cal D}+k$ to $\pgmv{clock}_q$,
verifying that $\pgmv{w}_q\geq k$ as a result.  To complete the lemma,
consider the first state $\delta$ where some $\pgmv{clock}_q$ has 
value $7{\cal D}+1$.  By the induction argument given in the 
proof of Lemma \ref{a8}, any \pgmv{clock} at distance $j$ 
from $\pgmv{clock}_q$ has had a value
of at least $7{\cal D}-j$ prior to state $\delta$.   
Therefore every \pgmv{clock} has contained a value of at least
$6{\cal D}+1$ prior to $\delta$, implying that each \pgmv{w} variable
is at least $3{\cal D}+1$ prior to $\delta$.  The state immediately
preceding $\delta$ thus satisfies proof obligation.        
\end{proofof}
\begin{proofof}{Lemma \ref{d2}}
by induction on $t$.  For $t=0$ let $\sigma'=\sigma$ to satisfy
the base case.  For $t>0$, we have
$\pgmv{clock}_q\leq 3t \;\wedge\; \pgmv{w}_q\leq 3t$ by hypothesis.
By the \textit{Echo} conditions of \pgmv{S2}, \pgmv{S3} and \pgmv{S6}, 
the \pgmv{clock} and \pgmv{w} values of $q$ remain at most $3t$ 
until all neighbors either \itp{i} complete cycles that observe these values
and write corresponding images to output registers
or \itp{ii} happen to have these values already in their output registers. 

Considering \itp{i}, for $r\in{\cal N}_q$ satisfying $\textit{dist}_{pr}=t+1$, 
the execution of \pgmv{S2} assures $\pgmv{w}_r\leq 3t+1$ within
one round, and $\pgmv{clock}_r$ is at most $3t+1$ if $r$ observes
no \textit{gap}, or assigned some value at most $w_r$ otherwise;
either case verifies the inductive hypothesis for $t+1$.
These considerations for \itp{i} also verify the second part of 
the lemma, which concerns a path of unperturbed processes, and
the same hypothesis with $3t$ replaced by $t$.  

Considering \itp{ii}, process $q$ may increment $\pgmv{clock}_q$ and 
$\pgmv{w}_q$ because $r\in{\cal N}_q$ happens already to have values
corresponding to $\pgmv{clock}_q$ and $\pgmv{w}_q$ in its
output register fields.  In this case, $q$ may increment its variables to 
at most $3t+1$ immediately.  Furthermore process $r$ 
may initially have its program counter at \pgmv{S7},
about to write its image variables in such a way that $q$ can observe
the \textit{cEcho} condition (even though $r$ would not actually read
and write in a full cycle).  Therefore, if $r$ executes \pgmv{S7}, process
$q$ can increment variables again to at most $3t+2$.  However, 
here a \textit{cEcho} condition will not be satisfied at $q$
until all neighbors complete full cycles, so $q$'s variables cannot
exceed $3t+2$ until $r$ completes a cycle.
When $r$ does complete a cycle, by the reasoning above for \itp{i} we deduce  
that $\pgmv{clock}_r\leq 3t+3$ and $\pgmv{w}_r\leq 3t+3$ for $r\in{\cal N}_q$.
\end{proofof}
\begin{proofof}{Lemma \ref{d1}}
Note that the lemma holds trivially if the initial state is 
$n$-perturbed.  For the case $k<n$ we use induction on $t$ and
nested induction on $k$ and suppose a based computation.  
For the base case $t=0$ consider $p\in{\cal N}_q$.
Since $q$ is perturbed, there is a path $P$ from $p$ to some perturbed $r$  
(possibly through $q$) of $k+2$ or fewer processes, which is not smooth.  
Because $\pgmv{clock}_p={\cal T}$, some neighboring pair of processes along
path $P$ has the property that one clock exceeds ${\cal T}-{\cal D}$
while the other is less than ${\cal T}-{\cal D}$.  Therefore some
process in path $P$ executes \pgmv{S4} in the first round.  By the 
arguments of Lemma \ref{d2} 
it follows that $p$ executes \pgmv{S4} within
$k+2$ rounds.   This completes the base case, but reasoning similar
to the nested induction also applies for $t>0$.  Finally, because
the initial state may not justify a based computation, two additional
rounds are added to conclude a $k+t+4$ bound.
\end{proofof}
\begin{proofof}{Lemma \ref{d3}}
Lemma \ref{d0} states that a process executes \pgmv{S4} at most
once in a computation, so it suffices to show that $p$ either
does not execute \pgmv{S4} or executes \pgmv{S4} within the
first $4+\min({\cal D},t+k)$ rounds.  If $p$ is unperturbed, Lemma \ref{d1}
implies the result.  If $p$ is perturbed, then for some perturbed 
region $P$ containing $p$, there is an unperturbed $q$ neighboring
some process of $P$ that executes \pgmv{S4} within the first 
$4+\min({\cal D},k)$ rounds by Lemma \ref{d1}.  Applying Lemma 
\ref{d2} we deduce that $\pgmv{clock}_p\leq 3\min({\cal D},k)$ holds after 
$\min({\cal D},k)$ additional rounds, and by arguments of Lemmas \ref{a8} and  
\ref{a9} process $p$ does not execute \pgmv{S4} in the remainder of
the computation.  Therefore, for perturbed $p$, the distance from
$p$ to a perturbed process is $t=0$ and after 
$4+\min({\cal D},k)+\min({\cal D},k)$ rounds, process $p$ does
not execute \pgmv{S4}.
\end{proofof} 
\begin{proofof}{Lemma \ref{d4}}
by induction on $t$.  The base case $t=0$ trivially follows from 
the domain of \pgmv{w} variables, which have non-negative values.
The same observation concerning the domain of \pgmv{w} variables 
simplifies the proof obligation to the case 
$\textit{dist}_{pq}\leq t$.  It is useful also to observe 
base cases for $t=1$ and $t=2$, since by the end of round two
the computation is based, which simplifies reasoning for higher
rounds.  For $t=1$ the verification is again trivial by the domain
of \pgmv{w} variables.  For $t=2$, it is required to show that
by the end of round two, $\pgmv{w}_p\geq 1$.  In fact any change
to $\pgmv{w}_p$ is an increase from its original value of zero,  
and at least one increment occurs because 
$\textit{wEcho}_p$ is observed by $p$ within two rounds
following $\sigma$.  No subsequent reduction to $\pgmv{w}_p$ 
results in a value less than one, since $\textit{wMin}_p$ is 
at least zero at all states.  This verifies the base case for $t=2$. 

Now suppose the hypothesis 
$\pgmv{w}_q\geq \lfloor(t-\textit{dist}_{pq})/2\rfloor$ for every $q$
such that $\textit{dist}_{pq}\leq t$ at some state $\sigma'$.
Note that no such process $q$ subsequently executes \pgmv{S4}
in the computation, by Lemma \ref{d2};  therefore any subsequent
change to $\pgmv{w}_q$ occurs by \pgmv{S2}.  If \pgmv{S2} assigns   
$\pgmv{w}_q$ a value at least $\lfloor((t+1)-\textit{dist}_{pq})/2\rfloor$
in the round following $\sigma'$, or if $\pgmv{w}_q$ already has such 
a value and does not decrease, then the induction step is verified. 
Therefore we consider the possibility that $\pgmv{w}_q$ either remains
unchanged or decreases below 
$\lfloor(t-\textit{dist}_{pq})/2\rfloor$ by execution of \pgmv{S2}.
A decrease only occurs if $\pgmv{w}_q>\textit{wMin}_q+1$, so a decrease 
below $\lfloor(t-\textit{dist}_{pq})/2\rfloor$ is only possible if
there is a neighbor $r\in{\cal N}_q$ satisfying 
$\pgmv{y}_q[r]\leq\lfloor(t-\textit{dist}_{pq})/2\rfloor-2$, which
would in turn imply that such a value existed in 
$\pgmv{w}_r$ in the previous round.  But by hypothesis,
$\pgmv{w}_r\geq\lfloor(t-\textit{dist}_{pr})/2\rfloor$, 
and since $r\in{\cal N}_q$ the value of $\pgmv{w}_r$ is at least 
$\lfloor(t-\textit{dist}_{pq}\pm 1)/2\rfloor$, 
which contradicts $\pgmv{y}_q[r]\leq\lfloor(t-\textit{dist}_{pq})/2\rfloor-2$. 
Therefore such a decrease to $\pgmv{w}_q$ cannot occur.  

The remaining case to consider is that 
$\pgmv{w}_q=\lfloor(t-\textit{dist}_{pq})/2\rfloor$ 
and does not change in the round following $\sigma'$.
Here there are two cases for $t$ and $q$, either 
$(t-\textit{dist}_{pq})$ is even or it is odd.  If 
$(t-\textit{dist}_{pq})$ is even, then 
$\lfloor(t-\textit{dist}_{pq})/2\rfloor$ is equal to
$\lfloor((t+1)-\textit{dist}_{pq})/2\rfloor$ and the 
hypothesis for $(t+1)$ is proved --- the value of $\pgmv{w}_q$
can remain unchanged in the round following $\sigma'$ and
satisfy the hypothesis.  If, however, $(t-\textit{dist}_{pq})$ is
odd, then $\pgmv{w}_q$ is required to increment to verify
the hypothesis for $(t+1)$.  Observe that if $(t-\textit{dist}_{pq})$ is
odd, then $\lfloor(t-\textit{dist}_{pq})/2\rfloor$ is equal to
$\lfloor((t-1)-\textit{dist}_{pq})/2\rfloor$, so we infer that
$\pgmv{w}_q=\lfloor(t-\textit{dist}_{pq})/2\rfloor$ held at
round $(t-1)$ (here we assume the hypothesis not only for $t$, 
but $(t-1)$ as well, which is permissible because base cases
for $t=1$ and $t=2$ have been verified).   Therefore by round
$(t+1)$, process $q$ observes $\textit{wEcho}_q$ and increments
$\pgmv{w}_q$, which verifies the hypothesis for $(t+1)$. 
\end{proofof}
\begin{proofof}{Lemma \ref{d5}}
by induction on $t$, for $t\geq 0$.
Note that round $t+2$ occurs in a based computation, 
since within two rounds following $\sigma$ the computation is based.  
The base case for induction is shown for $t=0$ and $t=1$, since
the main induction step relies on two previous rounds of a based computation.
For $t\leq 1$, since every \pgmv{clock} variable is at
least zero, the base cases are verified  
directly by the domain of \pgmv{clock} variables --- which are
at least zero at any state.  

Note that for any $t$,
$t-2\leq\textit{dist}_{pq}$ trivially satisfies the conclusion because
\pgmv{clock} variables are always at least zero;  therefore in the 
remainder of the proof we consider only the case of $q$ and $t$ satisfying 
$t-2>\textit{dist}_{pq}$.  
Now suppose the hypothesis holds for 
$t-1$ and $t-2$, $t\geq 2$, aiming to show that 
the hypothesis also holds for $t$, that is, that   
\pgmv{clock} variables satisfy the specified lower
bound by the end of round $t+2$.

By Lemmas \ref{d2} and \ref{d1}, by round $t$, any process in the set 
$R=\{\,r\;|\;\textit{dist}_{pr}\leq t-3\}$. 
has either executed \pgmv{S4} or will not do so throughout the 
remainder of the computation.  Therefore in round $t+2$, any reduction
to $\pgmv{clock}_r$ for $r\in R$ could only occur by \pgmv{S5}.  
Lemma \ref{d4} establishes that 
$\pgmv{w}_r\geq\lfloor((t+1)-\textit{dist}_{pr})/2\rfloor$ 
holds invariantly following round $t+1$.
So if process $r$ executes \pgmv{S5}, the result satisfies
$\pgmv{clock}_r\geq\lfloor((t-2)-\textit{dist}_{pr})/2\rfloor$, 
which would verify the inductive hypothesis for $r$ and round $t+2$.  
If $r$ does not execute \pgmv{S5} in round $x+1$, then consider
two cases for $r$.

\textbf{Case:} $t-\textit{dist}_{pr}$ is even.  Observe that
$\lfloor((t-2)-\textit{dist}_{pr})/2\rfloor$ differs from 
$\lfloor((t-3)-\textit{dist}_{pr})/2\rfloor$, meaning that
the obligation is to show that $\pgmv{clock}_r$ is either at least 
$\lfloor((t-2)-\textit{dist}_{pr})/2\rfloor$ by the end of 
round $t+1$, or that $\pgmv{clock}_r$ increments during round
$t+2$.  If the former holds, the hypothesis is proved, so suppose
$\pgmv{clock}_r=\lfloor((t-3)-\textit{dist}_{pr})/2\rfloor$ at
the end of round $t+1$.  Because $t-\textit{dist}_{pr}$ is even,
$\pgmv{clock}_r\geq\lfloor((t-4)-\textit{dist}_{pr})/2\rfloor$ by
hypothesis for $t-2$.  But this implies that during round $t+1$, 
the value of $\pgmv{clock}_r$ either did not change or was reduced
by \pgmv{S5}.  However a reduction by \pgmv{S5} would satisfy the 
hypothesis for $t$ as well, because of Lemma \ref{d4}'s bound on \pgmv{w}
variables.  The only remaining possibility is that $\pgmv{clock}_r$
does not change in round $t+1$, implying that $r$ observes 
\textit{cEcho} during round $t+2$.  Therefore, if 
$\pgmv{clock}_r\leq\textit{cMin}_r$ when $r$ observes \textit{cEcho}, 
then $\pgmv{clock}_r$ will increment either by \pgmv{S3} or \pgmv{S6}.
To show that $r$ does indeed observe \textit{cEcho}, we use the 
hypothesis for $t-1$ and each $q\in{\cal N}_r$.  If 
$\textit{dist}_{pq}\leq\textit{dist}_{pr}$, then by round $t+1$ (and 
throughout round $t+2$) the relation $\pgmv{clock}_q\geq\pgmv{clock}_r$
holds at least until $r$ increments its \pgmv{clock}.  
If $\textit{dist}_{pq}=\textit{dist}_{pr}+1$, then 
$\lfloor((t-2)-\textit{dist}_{pr})/2\rfloor$ and 
$\lfloor((t-2)-\textit{dist}_{pq})/2\rfloor$ are equal, 
and again the relation $\pgmv{clock}_q\geq\pgmv{clock}_r$
holds at least until $r$ increments its \pgmv{clock}.  

\textbf{Case:} $t-\textit{dist}_{pr}$ is odd.  A similar detailed
argument can be given for this case, but there is a simpler approach:
$\lfloor((t-2)-\textit{dist}_{pr})/2\rfloor$ and 
$\lfloor((t-3)-\textit{dist}_{pr})/2\rfloor$ are equal,
so the hypothesis for $t-1$ and $r$ directly suffice to verify
the hypothesis for $t$. 
\end{proofof}
\begin{proofof}{Lemma \ref{e1}}
In any computation, either some process executes 
a double-reset or no process does so.  
In the latter case, the core system component stabilizes within 
$\cal M$ rounds, and the repair timer concurrently reaches the 
timer-final condition within $O({\cal T})$ rounds by Theorem \ref{stab}.
This demonstrates ${\cal M}+O({\cal T})$ stabilization time if
no double-reset occurs;  the 
same argument applies to the case where any  
double-reset occurs by \pgmv{S4}
and not by the core system.  Lemma \ref{d0} implies that 
a double-reset occurs at most once 
for each process in this case. 

Now consider the possibility that the core system
executes a double-reset at least once 
in a computation.  All such assignments cease after the base
system stabilizes, which occurs within $\cal M$ rounds, so the 
system stabilization time is ${\cal M}+O({\cal T})$.  To  
show that any process executes 
a double-reset at most once,
we demonstrate that the core system stabilizes before
$\pgmv{clock}>{\cal T}-{\cal D}$ holds at any process,
since Requirement \ref{interface} prevents repeated resets
of the \pgmv{clock} so long as $\pgmv{clock}\leq{\cal T}-{\cal D}$. 

If any double-reset assignment occurs, then
within $\cal D$ rounds thereafter, a state $\sigma$ occurs such that 
each \pgmv{clock} is at most $3{\cal D}$ by Lemma \ref{d2}, and
also within $\cal D$ rounds, time-accuracy holds and is 
invariant thereafter by Theorem \ref{d6}.  Although Theorem \ref{d6}
is conditioned on $k<n$ for a $k$-perturbed initial state, its
proof arguments are valid for the case of an $n$-faulty initial
state, provided some process executes a double-reset
in the first round.  While we do not suppose that a double-reset 
occurs in the first round, the state preceding the first double-reset
can be considered as the initial state for the
subsequent computation, so that Theorem \ref{d6}'s results apply  
for the suffix computation.
Time accuracy for the extreme case of an $n$-faulty initial
state implies for $\pgmv{clock}=t$ that at least 
$(t-{\cal D}-5{\cal D})=t-6{\cal D}$ rounds have transpired.
Therefore, if ${\cal T}-{\cal D}\geq X+6{\cal D}$,   
where $X$ is the number of rounds needed for stabilization,
then as soon as time accuracy holds, 
no \pgmv{clock} increases beyond
${\cal T}-{\cal D}$ until the core system has stabilized.  Requirement
\ref{core-converge} implies stabilization within 
${\cal T}-7{\cal D}$ rounds, which ensures
that the core system stabilizes before there is the possibility of
a second double-reset.  To complete the proof we address the period
between the first double-reset and before time accuracy holds.  This
is at most $\cal D$ rounds, and it is easy to show that no \pgmv{clock}
increases from zero to beyond ${\cal T}-{\cal D}$ within $\cal D$ 
rounds, so a second double-reset does not occur in the period before 
time accuracy holds.  
\end{proofof}

\end{document}